\documentclass[12pt]{article}

\usepackage{geometry} 
\usepackage{graphicx} 
\usepackage[colorlinks,linkcolor=blue,anchorcolor=blue,citecolor=blue]{hyperref}
\usepackage{float} 
\usepackage{wrapfig} 
\usepackage{lipsum} 
\usepackage{exscale}
\usepackage{relsize}
\usepackage{indentfirst}
\usepackage{amssymb}
\usepackage{tikz}
\usepackage[inline]{enumitem}

\usepackage{makecell}
\usepackage{amsthm}
\usepackage{amsmath,bm}
\usepackage{abstract}
\usepackage{setspace}
\usepackage[figuresright]{rotating}
\usepackage{threeparttable}
\usepackage[misc]{ifsym}

\usetikzlibrary{shapes.geometric, arrows}

\DeclareMathOperator{\Tr}{Tr}
\DeclareMathOperator{\Hull}{Hull}
\DeclareMathOperator{\diag}{diag}
\DeclareMathOperator{\LCD}{LCD}
\DeclareMathOperator{\LCDoe}{LCD_{o,e}}
\DeclareMathOperator{\LCDeo}{LCD_{e,o}}
\DeclareMathOperator{\LCDoo}{LCD_{o,o}}
\DeclareMathOperator{\LCDee}{LCD_{e,e}}

\renewcommand{\arraystretch}{1.8}

\numberwithin{equation}{section}

\newtheorem{thm}{Theorem}[section]
\newtheorem{defn}[thm]{Definition}
\newtheorem{lem}[thm]{Lemma}
\newtheorem{coro}[thm]{Corollary}
\newtheorem{prop}[thm]{Proposition}
\newtheorem{remark}[thm]{Remark}
\newtheorem{exam}[thm]{Example}

\begin{document}
\title{New Constructions of Optimal Binary LCD Codes\let\thefootnote\relax\footnotetext{
 E-Mail addresses: wanggd@163.com (G. Wang),\\ shengweiliu@mails.ccnu.edu.cn (S. Liu), $^{\ast}$hwliu@ccnu.edu.cn (H. Liu).}}

\author{~Guodong Wang,~Shengwei Liu,~Hongwei Liu$^{\ast}$}

\date{\small School of Mathematics and Statistics, Central China Normal University, Wuhan, 430079, China \\}
\maketitle

\begin{abstract}
    Linear complementary dual (LCD) codes provide an optimum linear coding solution for the two-user binary adder channel.
LCD codes also can be used against side-channel attacks and fault non-invasive attacks.
In this paper, we obtain a lower bound on the distance of binary LCD codes through expanded codes.
We give necessary and sufficient conditions to extend binary $[n,k]$ LCD codes to binary $[n+1,k]$ and $[n+1,k+1]$ LCD codes.
A sufficient condition for a binary $[n,k,d-1]$ LCD code constructed from a binary $[n+1,k,d]$ $\LCDoe$ code is also given.
Finally, we construct some new binary LCD codes by using our LCD bounds and some methods such as puncturing, shortening, expanding and extension.
In particular, we improve some previously known range of $d_{\LCD}(n, k)$ of lengths $38 \le n \le 40$ and dimensions $9 \le k \le 15$.
We also obtain some values or range of $d_{\LCD}(n, k)$ with $41 \le n \le 50$ and $6 \le k \le n-6$.

\noindent \textbf{Keywords:}
LCD codes, optimal binary LCD codes, expanded codes, lower bounds for binary LCD codes.

\noindent \textbf{2020 Mathematics Subject Classification:} 94B15, 94B60
\end{abstract}

\section[Introduction]{Introduction}
An $[n,k]$ linear code $C$ over the finite field $\mathbb{F}_q$ is called {\em linear complementary dual} (LCD for short) if $C \cap C^{\bot} = \{\bm{0}_n\}$, where $\bm{0}_n$ denotes the zero vector of length $n$.
Massey \cite{MA92} introduced the notion of LCD codes and gave an optimum linear coding solution of two linear binary LCD codes for the two-user binary adder channel and showed that LCD codes are asymptotically good.
Carlet and Guilley \cite{CG16} gave applications of LCD codes in side-channel attacks and fault non-invasive attacks.
Some new LCD codes were constructed from existing LCD codes by using methods such as: direct sum, direct product, puncturing, shortening, extension and matrix product (see \cite{BS21}, \cite{CG16}, \cite{CMT19}, \cite{LLY20}).

We denote
\begin{align*}
    d(n,k)_{\mathbb{F}_q} &= \max\{d(C) : C \ \text{is an $[n,k,d(C)]$ linear code over $\mathbb{F}_q$}\},\\
    d_{\LCD}(n,k)_{\mathbb{F}_q} &= \max\{d(C) : C \ \text{is an $[n,k,d(C)]$ LCD code over $\mathbb{F}_q$}\}.
\end{align*}
For convenience, we abbreviate $d(n,k)_{\mathbb{F}_2}$ and $d_{\LCD}(n,k)_{\mathbb{F}_2}$ as $d(n,k)$ and $d_{\LCD}(n,k)$.
We call a linear code with parameters $[n, k, d(n,k)_{\mathbb{F}_q}]$ as an {\em optimal} code.
The determination of $d(n,k)_{\mathbb{F}_q}$ and the search for corresponding optimal codes are ongoing and meaningful pursuits,
and some results can be found in codetables \cite{GTB22}.
An LCD code with parameters $[n, k, d_{\LCD}(n,k)_{\mathbb{F}_q}]$ is called an {\em optimal} LCD code.

Two codes $C$ and $C'$ are {\em equivalent} if one can be obtained from the other by
permuting the coordinates.
If the generator matrix of one code can be obtained by multiplying the generator matrix of the other code by a monomial matrix,
then the two codes are called {\em monomial equivalent}.
Carlet et al. \cite{CMT18} showed that for $q \ge 4$, any linear code over $\mathbb{F}_q$ is monomial equivalent to some LCD code.
We know that monomial equivalence maintains Hamming distance, which means that LCD codes over $\mathbb{F}_q(q \ge 4)$ can achieve the same parameters as linear codes.
When $q=2$ or $q=3$, determining the value of $d(n,k)_{\mathbb{F}_q}$ and finding the corresponding optimal LCD codes are also meaningful and ongoing works.
The primary objective of this article is to investigate the maximum possible minimum distance of binary LCD codes with specific lengths $n$ and dimensions $k$.

In \cite{GKL18}, \cite{HS19}, \cite{AH20}, \cite{BS21}, \cite{IS21} and \cite{LSH2022} the values of $d_{\LCD}(n,k)$ have been completely determined for $n \le 24$,
and the values of $d_{\LCD}(n,k)$ for $n \le $ 30 were settled completely except for the values of $d_{\LCD}(26,15)$, $d_{\LCD}(28,17)$, $d_{\LCD}(30,11)$ and $d_{\LCD}(30,19)$.
For $31 \le n \le 40$, about half of the values of $d_{\LCD}(n,k)$ have been obtained, and some classification results on LCD codes with such lengths were also presented (\cite{AH20}, \cite{BS21}, \cite{IS21}, \cite{LSH2022}).
When $k \le 5$, there are corresponding formulae available to calculate the values of $d_{\LCD}(n,k)$ (\cite{AH20}, \cite{AH21}, \cite{AHS21}, \cite{DKO17}, \cite{GKL18}, \cite{HS19}, \cite{LYL2022}).

Expanded codes can be used to study Reed-Solomon codes and cyclic codes (\cite{RE91}, \cite{SK96}, \cite{STK89}, \cite{WU11}).
MacWilliams \cite{MAC70} and Seguin \cite{SE95} characterized that under certain conditions an expanded cyclic code remains cyclic.
When $m \ge 2$, a linear code over $\mathbb{F}_{2^m}$ is monomial equivalent to an LCD code as described in \cite{CMT18}.
This implies that, we can obtain some optimal LCD codes over $\mathbb{F}_{2^m}$ by transforming the best known linear codes in codetables \cite{GTB22} using monomial equivalent.
By expanding these optimal LCD codes over $\mathbb{F}_{2^m}$ with respect to a self-dual basis, we can obtain some binary LCD codes.

On the other hand, puncturing and extension are also very useful methods to construct LCD codes.
The punctured code of an $[n, k, d] (d \ge 2)$ $\LCDoe$ code by puncturing one coordinate position can not be LCD \cite{BS21}, where $\LCDoe$ denotes an odd-like binary LCD code with even-like dual.
In \cite{BS21} and \cite{LSH2022}, the authors employed extension methods to construct longer binary LCD codes from shorter binary LCD codes.
In this paper, we use extension methods to construct binary LCD codes from either binary LCD or linear codes.
Furthermore, we provide a sufficient condition for constructing a binary $[n,k,d-1]$ LCD code from a binary $[n+1,k,d]$ $\LCDoe$ code.

The paper is organized as follows.
In Section 2, we present some characterizations on LCD codes and some known methods for constructing binary LCD codes.
In Section 3, we give a lower bound for the minimum distance of binary LCD codes by expanded codes.
Then we provide a necessary and sufficient condition for extending binary $[n, k]$ LCD codes to binary $[n+1, k]$ and $[n+1, k+1]$ LCD codes.
We also introduce a method for extending binary $[n, k]$ linear codes with $1$-dimension hull to binary $[n+1, k]$ and $[n+1, k+1]$ LCD codes.
In Section 4, we give a sufficient condition for constructing a binary $[n,k,d-1]$ LCD code from a binary $[n+1,k,d]$ $\LCDoe$ code, where $d \ge 3$ and $k \ge 2$.
In Section 5, we construct some new binary LCD codes by using our LCD bounds and some methods such as puncturing, shortening, expanding and extending codes.
We improve some previously known range of $d_{\LCD}(n, k)$ of lengths $38 \le n \le 40$ and dimensions $9 \le k \le 15$.
We also determine some values of $d_{\LCD}(n, k)$ with $41 \le n \le 50$ and $6 \le k \le n-6$.
If the exact value of $d_{\LCD}(n, k)$ is not obtained, we give an interval of $d_{\LCD}(n, k)$.
Tables 1-4 show our new constructed LCD codes and methods we used.
Tables 5-6 provide the values or the range of $d_{\LCD}(n,k)$. All computations in this paper are performed in MAGMA \cite{BC97}.

\section{Preliminaries}

Let $\mathbb{F}_q$ be the finite field of order $q$, where $q$ is a prime power. Let $\mathbb{F}_q^n$ be the $n$-dimensional vector space over $\mathbb{F}_q$.
The (Hamming) {\em distance} $d(\bm{x}, \bm{y})$ between two vectors $\bm{x}, \bm{y} \in \mathbb{F}_q^n$ is the number of coordinate
positions in which they differ. The (Hamming) {\em weight} $wt(\bm{x})$ of a vector $\bm{x} \in \mathbb{F}_q^n$ is the
number of its nonzero coordinates. For any $\bm{x}=(x_1,...,x_n),\bm{y}=(y_1,...,y_n) \in \mathbb{F}_q^n$, the
{\em Euclidean (standard) inner product} of $\bm{x}$ and $\bm{y}$ is defined by $\langle \bm{x},\bm{y} \rangle=\sum_{i=1}^nx_iy_i$.

An $[n,k, d]$ linear code $C$ is a $k$-dimensional subspace of
the vector space $\mathbb{F}_q^n$, where $d$ is the smallest weight among all nonzero codewords of $C$, which is called
the {\em minimum weight} (or {\em minimum distance}) of the code.
The {\em dual} code $C^{\bot}$ of $C$ is defined as $C^{\bot} = \{\bm{x} \in \mathbb{F}_q^n : \langle \bm{x},\bm{y} \rangle =
0$ for all $\bm{y} \in C\}$. For convenience, we denote by $d^{\bot}$ the minimum distance of $C^{\bot}$. The {\em Hull} of $C$ is defined as $\Hull(C)=C \cap C^{\bot}$.
A $k \times n$ matrix $G$ whose rows form a basis of $C$ is called a {\em generator matrix} of this code,
and a generator matrix $H$ of $C^{\bot}$ is called a {\em parity-check matrix} of $C$.

An $[n,k]$ linear code $C$ over $\mathbb{F}_q$ is called {\em self-orthogonal} if $C \subseteq  C^{\bot}$ and {\em self-dual} if $C = C^{\bot}$.
A linear code $C$ is called {\em linear complementary dual} (LCD for short) if $C \cap C^{\bot} = \{\bm{0}_n\}$, where $\bm{0}_n$ denotes the zero vector of length $n$.
In particular, codes over $\mathbb{F}_2$ are called {\em binary codes}. A binary code is called {\em even-like} if the weights of all codewords are even, and {\em odd-like} otherwise.
An odd-like binary LCD code with odd-like dual is denoted by {\em $\LCDoo$}.
The notion of $\LCDeo$, $\LCDoe$ and $\LCDee$ can be defined similarly.
Obviously there is no binary even-like LCD code with even-like dual and hence there is no $\LCDee$ code.

Let $C$ be an $[n,k]$ code, $\mathcal{N}$ be a set of $|\mathcal{N}|$ coordinates, where $|\mathcal{N}|$ denotes the cardinality of $\mathcal{N}$.
The {\em punctured code} of $C$ on $\mathcal{N}$ is the code obtained by deleting all the coordinates in $\mathcal{N}$ for each codeword of $C$.
Let
\[
    C(\mathcal{N})=\{ \bm{c} \in C : \text{the coordinate of $\bm{c}$ are $0$ on $\mathcal{N}$}\}
\]
be the subcode of $C$.
The puncturing code of $C(\mathcal{N})$ on $\mathcal{N}$ is called a {\em shortened code} of $C$ on $\mathcal{N}$.

Throughout this paper, let $I_k$ be the identity matrix of order $k$, $E_k$ be the all-one matrix of order $k$, and
\setlength{\arraycolsep}{1pt}
\renewcommand\arraystretch{0.8}
$J_2=
\begin{pmatrix}
    0 & 1 \\
    1 & 0
\end{pmatrix}$.
Let $G$ be a matrix over $\mathbb{F}_q$ and $G^T$ be the transpose matrix of $G$.
Denote by  $\diag(A_1,...,A_s)$ the block diagonal matrix whose elements on the diagonal are matrices $A_1,...,A_s$.
Let $GL_k(\mathbb{F}_2)$ be the general linear group of order $k$ over $\mathbb{F}_2$.
If $\bm{c}_1,...,\bm{c}_k \in \mathbb{F}_2^n$, we define $\langle \bm{c}_1,...,\bm{c}_k\rangle_{\mathbb{F}_2}$ to be the linear code generated by $\{\bm{c}_1,...,\bm{c}_k\}$.

The following characterization on LCD codes was given by Massey \cite{MA92}.

\begin{lem}\label{malcd}
    Let $G$ and $H$ be a generator matrix and a parity-check matrix of a code $C$ over $\mathbb{F}_q$,
respectively. Then the following are equivalent:
\begin{enumerate*}[label =(\arabic*)]
\item $C$ is LCD.
\item $C^{\bot}$ is LCD.
\item $GG^T$ is nonsingular.
\item $HH^T$ is nonsingular.
\end{enumerate*}
\end{lem}

\begin{defn}\label{defibas}
Let $C$ be an $[n,k]$ linear code over $\mathbb{F}_q$. Then

\noindent (1)~A basis $\{\bm{c}_1, \bm{c}_2,..., \bm{c}_k\}$ of $C$ is called orthonormal if for any $i, j \in \{1, 2,..., k\}$,
$\langle \bm{c}_i,\bm{c}_j \rangle=\delta_{i,j}$, where $\delta_{i, j}$ is the Kronecker symbol.

\noindent (2)~If $k$ is even, a basis $\{\bm{b}_1, \bm{b}_1',..., \bm{b}_{k/2}, \bm{b}_{k/2}'\}$ of $C$ is
called symplectic when the following three conditions are satisfied:
(i). $\langle \bm{b}_i,\bm{b}_i \rangle = \langle \bm{b}_i' , \bm{b}_i' \rangle = 0$; (ii). $\langle \bm{b}_i , \bm{b}_j \rangle = \langle \bm{b}_i , \bm{b}_j' \rangle = 0$ if $i \neq j$; (iii). $\langle \bm{b}_i , \bm{b}_i' \rangle = 1$.
\end{defn}
From now on, all codes are assumed to be binary in this section.

\begin{lem}\cite{LWL2019}\label{carleoo}
Let $C$ be an $[n,k]$ code with $\dim(\Hull(C))=s$. Then there exists a generator matrix $G$ of $C$ such that
    \setlength{\arraycolsep}{1pt}
    \renewcommand\arraystretch{0.8}
    \[
        GG^T=
        \begin{pmatrix}
            0_s & 0 \\
            0 & A
        \end{pmatrix}, \ \text{where}\
        A=\begin{cases}
            \diag(J_2,...,J_2)_{\frac{k-s}{2}},  & \text{if $C$ is even-like}, \\
            I_{k-s},  & \text{if $C$ is odd-like}.
        \end{cases}
    \]
\end{lem}

\begin{lem}\cite{CMT19}\label{carletob}
   Let $C$ be an $[n,k]$ $\LCD$ code. Then

\noindent (1) $C$ is an odd-like $\LCD$ code if and only if $C$ has an orthonormal basis.

\noindent (2) $C$ is an even-like $\LCD$ code if and only if $k$ is even and $C$ has a symplectic basis.
\end{lem}

\begin{lem}\label{impp}
    Let $C$ be a binary odd-like $[n,k]$ $\LCD$ code, $\bm{c} \in C$ with odd weight.
Then $\bm{c}$ can be extended to a basis $\{\bm{g}_1=\bm{c},\bm{g}_2,...,\bm{g}_k\}$ of $C$ such that $\langle \bm{g}_1, \bm{g}_i \rangle = 0$ for $i=2,...,k$.
\begin{proof}
There exist $\bm{g}_2',...,\bm{g}_k' \in C$ such that $\{\bm{g}_1=\bm{c},\bm{g}_2',...,\bm{g}_k'\}$ is a basis of $C$.
Let $\bm{g}_i = \bm{g}_i' + \langle \bm{g}_1, \bm{g}_i' \rangle \bm{g}_1$, $i=2,...,k$. Then $\langle \bm{g}_1,\bm{g}_i \rangle=\langle \bm{g}_1,\bm{g}_i' \rangle + \langle \bm{g}_1, \bm{g}_i' \rangle \langle \bm{g}_1,\bm{g}_1 \rangle=0$, $i=2,...,k$.
Obviously, $\{\bm{g}_1,\bm{g}_2,...,\bm{g}_k\}$ is a basis of $C$.
\end{proof}
\end{lem}

It is easy to know that if $C$ is an $[n,k]$ LCD code, then $\bm{1}_n \in C $ if and only if $C^{\bot}$ is even-like, i.e. $\bm{1}_n \in C$ if and only if $C$ is an $\LCDoe$ code.

\begin{lem}\label{allonev}
    If $C$ is a binary $[n,k]$ $\LCDoe$ code,
        \setlength{\arraycolsep}{1pt}
        \renewcommand\arraystretch{0.8}
        $G=\begin{pmatrix}
        \bm{g}_1  \\
        \vdots \\
        \bm{g}_k  \end{pmatrix}$ is a generator matrix of $C$ with $GG^T=I_k$,
then $\bm{g}_1+\bm{g}_2+\cdots+\bm{g}_k=\bm{1}_n$.
\begin{proof}
    Since $C$ is $\LCDoe$, then $\bm{1}_n \in C$,
    suppose $\bm{1}_n = \lambda_1\bm{g}_1+\cdots+\lambda_k\bm{g}_k$ where $\lambda_i \in \mathbb{F}_2$ for $i=1,...,k$.
    Since $\langle \bm{1}_n,\bm{g}_i \rangle = \lambda_i\langle \bm{g}_i,\bm{g}_i \rangle=\lambda_i=1$, for $i=1,...,k$ and $\{\bm{g}_1,...,\bm{g}_k\}$ is an orthonormal basis,
    we have $\bm{g}_1+\bm{g}_2+\cdots+\bm{g}_k=\bm{1}_n$.
\end{proof}
\end{lem}

It is a very useful method to construct binary LCD codes by puncturing and shortening known LCD codes.
This will be used later in the construction of new binary LCD codes in Section 5.
Many new LCD codes in this paper are obtained by such methods.
Bouyuklieva \cite{BS21} first studied the properties of puncturing and shortening operations for binary LCD codes,
and thus constructed many optimal binary LCD codes.
The following lemma can be easily derived from Corollaries 2 and 6 in Reference \cite{BS21}.

\begin{lem} \label{pross3}
    Let $C$ be an $[n, k, d]$ $\LCD$ code with odd $d$.

\noindent(1)~If $k$ is even and $\{\bm{c}_1,..., \bm{c}_k\}$ is an orthogonal basis of $C$,
then the vectors $\{(1, \bm{c}_1), ... , (1, \bm{c}_k)\}$ generate an even-like $[n+1, k, d+1]$ $\LCD$ code.

\noindent(2)~If $k$ is odd and $\{\bm{c}_1,..., \bm{c}_k\}$ is an orthogonal basis of $C$,
then the vectors $\{(1, 1, \bm{c}_1), ... , (1, 1, \bm{c}_k)\}$ generate an old-like $[n+2, k, \ge d+1]$ $\LCD$ code.
\end{lem}

\begin{lem}\cite{BS21}\label{lema2}
    Let $C$ be an $[n, k, d](d \ge 2)$ $\LCD$ code and $d^{\bot} \ge 2$. Then exactly one of shortened code and
    punctured code on a coordinate position is $\LCD$.
\end{lem}

\begin{lem}\cite{BS21}\label{pro33}
    Let $C$ be an even-like $[n,k, d]$ LCD code. Then the punctured code of $C$ on any one
coordinate is again $\LCD$.
\end{lem}

\begin{lem}\cite{BS21}\label{pro34}
    Let $C$ be an odd-like $[n,k, d]$ $\LCD$ code.

\noindent(1)~If $\bm{1}_n \in C$,
then the shortened code of $C$ on any one coordinate is again $\LCD$.

\noindent(2)~If $\bm{1}_n \notin C$,
then there are integers $i, j \in \{1,...,n\}$ such that the shortened code of $C$
on the $i$th coordinate and the punctured code of $C$ on the $j$th coordinate are $\LCD$ codes.
\end{lem}

LCD codes can also be constructed by performing puncturing or shortening operations on multiple coordinates for linear codes.
Li et al.\cite{LSH2022} first introduced this method and used it to construct many binary LCD codes with good parameters.

\begin{prop}\cite{LSH2022}\label{pro35}
Let $C$ be an $[n, k, d]$ linear code with $\dim(\Hull(C))=s$. Then
there exists a set $T$ of $s$ coordinates such that the shortened code of $C$ on coordinates in $T$ is an
$[n-s, k-s, \ge d]$ $\LCD$ code.
\end{prop}

\begin{prop}\cite{LSH2022}\label{pro36}
Let $C$ be an $[n, k, d]$ linear code with $\dim(\Hull(C))=s$. If $s < d$, then
there exists a set $T$ of $s$ coordinates such that the punctured code of $C$ on $T$
is an $[n - s, k, \ge d -s]$ $\LCD$ code.
\end{prop}

\section{Two methods for constructing LCD codes}
In this section, we provide two methods to construct LCD codes.
By using expansion methods, we obtain a distance lower bound for binary LCD codes.
And we also give some characterization for extending a shorter binary LCD or not LCD code to an longer LCD code.
In Section 5, we will use these two methods to construct binary LCD codes with good parameters.

\subsection{Construction of LCD codes with expansion methods}

Let $\mathbb{F}_{q^m}$ be the finite extension of $\mathbb{F}_q$, the field $\mathbb{F}_{q^m}$ can be viewed as
an $\mathbb{F}_q$-vector space of dimension $m$. Let $\{\alpha_1,...,\alpha_m\}$ be a basis of $\mathbb{F}_{q^m}$ over $\mathbb{F}_q$.
Any vector $x$ of $\mathbb{F}_{q^m}$ can be uniquely written as $x=a_1\alpha_1+a_2\alpha_2+\cdots+a_m\alpha_m$. Then we
have the following $\mathbb{F}_q$-linear isomorphism:
\begin{align*}
    \phi:\mathbb{F}_{q^m} &\longrightarrow \mathbb{F}^m_q,\\
    x=a_1\alpha_1+\cdots+a_m\alpha_m &\longmapsto (a_1,a_2,...,a_m).
\end{align*}
This isomorphism can be extended as follows:
\begin{align*}
    \phi_n:\mathbb{F}_{q^m}^n &\longrightarrow \mathbb{F}^{mn}_q,\\
    (x_1,x_2,...,x_n) &\longmapsto (\phi(x_1),\phi(x_2),...,\phi(x_n)).
\end{align*}

\begin{defn}
    Let $\mathbb{F}_{q^m}$ be the finite extension field of $\mathbb{F}_q$, and let $C$ be an $[n,k]$ linear code over $\mathbb{F}_{q^m}$.
Let $\{\alpha_1,...,\alpha_m\}$ be a basis of $\mathbb{F}_{q^m}$ over $\mathbb{F}_q$.
The expanded code $\Phi_n(C)$ of $C$ with respect to the basis $\{\alpha_1,...,\alpha_m\}$ over $\mathbb{F}_q$ is defined as
\[
    \Phi_n(C)=\{\phi_n(\bm{c}):\bm{c} \in C\},
\]
where $\phi_n$ is the $\mathbb{F}_q$-linear isomorphism defined as above.
\end{defn}

The notions and results in the following can be found in \cite{FFR}.

Let $\mathbb{F}_{q^m}$ be the finite extension field of $\mathbb{F}_q$. The {\em trace} function $\Tr_{\mathbb{F}_{q^m}/\mathbb{F}_q}(\cdot)$ (denoted by $\Tr(\cdot)$ for short)
over $\mathbb{F}_q$ is defined by
\[
    \Tr(x)=x+x^q+\cdots+x^{q^{m-1}}, \forall \, x \in \mathbb{F}_{q^m}.
\]

A basis $\{\alpha_1,...,\alpha_m\}$ of $\mathbb{F}_{q^m}$ over $\mathbb{F}_q$ is called {\em self-dual}, if it satisfies
\[
    \Tr(\alpha_i \alpha_j)=\begin{cases}
        1,  & \text{for}\ i = j, \\
        0,  & \text{for}\ i \neq j.
    \end{cases}( 1 \le i,j \le m)
\]

If there exists a self-dual basis $\{\alpha_1,...,\alpha_m\}$ of $\mathbb{F}_{q^m}$ over $\mathbb{F}_q$, then for any $a,b \in \mathbb{F}_{q^m}$, we have
\begin{align}\label{equfa1}
    a &= \Tr(a\alpha_1)\alpha_1+\cdots+\Tr(a\alpha_m)\alpha_m, \\ \label{equfa2}
    \Tr(ab) &=\Tr(\alpha_1a)\Tr(\alpha_1b)+\cdots+\Tr(\alpha_ma)\Tr(\alpha_mb).
\end{align}

The following results show that some algebraic properties of linear codes are invariant under the map $\Phi_n$.

\begin{thm}\label{exco}
    Let $\mathbb{F}_{q^m}$ be a finite extension field of $\mathbb{F}_q$ with a self-dual basis $\{\alpha_1, ...,\alpha_m\}$ and
let $C$ be an $[n,k, d]$ linear code over $\mathbb{F}_{q^m}$. Let $\Phi_n(C)$ be the expanded code of $C$ with respect to the basis $\{\alpha_1, ...,\alpha_m\}$.
Then we have

\noindent (1) $\Phi_n(C)$ is an $[mn, mk, \ge d]$ linear code over $\mathbb{F}_q$.

\noindent (2) $\Hull(\Phi_n(C))$ = $\Phi_n(\Hull(C))$.

\noindent (3) $\Phi_n(C)$ is $\LCD$ if and only if $C$ is $\LCD$.
\begin{proof}
    We will only give a proof for $(2)$, the proofs of others are easy to obtain.
    For any $\bm{x}=(x_1,...,x_n) \in C$, $\bm{y}=(y_1,...,y_n) \in C^{\bot}$, by Equations (\ref{equfa1}) and (\ref{equfa2}), we have
\begin{align*}
    \langle \phi_n(\bm{x}),\phi_n(\bm{y}) \rangle &= \sum_{i=1}^n \langle \phi(x_i),\phi(y_i) \rangle \\
    & = \sum_{i=1}^n \langle (\Tr(x_i\alpha_1),...,\Tr(x_i\alpha_m)), (\Tr(y_i\alpha_1),...,\Tr(y_i\alpha_m)) \rangle \\
    & = \sum_{i=1}^n \sum_{j=1}^m \Tr(x_i\alpha_j) \Tr(y_i\alpha_j) \\
    & = \sum_{i=1}^n \Tr(x_iy_i)= \Tr(\langle \bm{x},\bm{y} \rangle)  = 0.
\end{align*}
We conclude that $\Phi_n(\Hull(C)) \subseteq \Hull(\Phi_n(C))$.

Let $\bm{u} \in C \backslash \Hull(C)$, there exists $\bm{v} \in C$ such that $\langle \bm{u},\bm{v} \rangle \neq 0$.
Let $0 \neq \lambda \in \mathbb{F}_{q^m}$, we have $\langle \bm{u},\lambda\bm{v} \rangle = \lambda \langle \bm{u},\bm{v} \rangle \neq 0$.
There exists an element $0 \neq \lambda_0 \in \mathbb{F}_{q^m}$ such that $\langle \phi_n(\bm{u}),\phi_n(\lambda_0\bm{v}) \rangle=\Tr(\langle \bm{u},\lambda_0\bm{v} \rangle)=\Tr(\lambda_0\langle \bm{u},\bm{v} \rangle) \neq 0$ since the trace map $\Tr$ is not a zero map.
Then $\phi_n(\bm{u}) \notin \Hull(\Phi_n(C))$ and $\Phi_n(\Hull(C)) = \Hull(\Phi_n(C))$.
The result then follows.
\end{proof}
\end{thm}

Note that self-dual bases always exist in finite fields with even characteristic \cite{JMV90}.
When $m \ge 2$, a linear code over $\mathbb{F}_{2^m}$ is monomial equivalent to some LCD code \cite{CMT18}.
We can obtain optimal LCD codes over $\mathbb{F}_{2^m}$ via monomial equivalent. Then,
by expanding these optimal LCD codes over $\mathbb{F}_{2^m}$ with respect to a self-dual basis,
we can obtain some binary LCD codes.

In \cite{CMT19}, the authors demonstrated that $d_{\LCD}(n,k-1) \ge d_{\LCD}(n,k)$.
By combining this result with the simple inequality $d_{\LCD}(n+1,k) \ge d_{\LCD}(n,k)$,
it is easy to deduce the following conclusion.

\begin{prop}\label{lbound}
Assume the notations are given above. Then
\[
    d_{\LCD}(n,k) \ge d(\lfloor  n/m \rfloor , \lceil k/m \rceil)_{\mathbb{F}_{2^m}},
\]
where $\lfloor x \rfloor$ denotes the greatest integer $\le x$, and
$\lceil x \rceil$ denotes the least integer $\ge x$.

\end{prop}

\begin{remark}
    From the codetables \cite{GTB22}, we know $d(21,18)_{\mathbb{F}_4}=3$ and $d(42,36)=3$.
With this bound, we have $d_{\LCD}(42,36) = 3$, which mean this bound is tight.
\end{remark}

\subsection{Construction of LCD codes with extension methods}
In this section, all codes are assumed to be binary.
In \cite{LSH2022}, the authors constructed $[n+1,k+1]$ LCD codes by $[n,k]$ LCD codes.

\begin{prop}\cite{LSH2022}\label{pro37}
    Let $C$ be an $[n, k]$ $\LCD$ code with a generator matrix $G$. Let $\bm{x} \in C^{\bot}$ and $C(\bm{x})$ be the linear code with the generator matrix
    \setlength{\arraycolsep}{1pt}
    \renewcommand\arraystretch{0.8}
    \begin{equation}\label{equ2}
        G(\bm{x})=
        \begin{pmatrix}
            \bm{x} & 1 \\
            G & 0\\
        \end{pmatrix}.
    \end{equation}
    If $wt(\bm{x})$ is even, then $C(\bm{x})$ is a binary $[n+1, k+1]$ $\LCD$ code.
\end{prop}

When an $[n, k]$ LCD code is extended to an $[n+1, k]$ LCD code,
its dual code can be extended to an $[n+1, n-k+1]$ LCD code from an $[n, n-k]$ LCD code.
Based on this basic result, we give some other characterization for extending an $[n,k]$ LCD code to either an $[n+1,k]$ or $[n+1,k+1]$ LCD code.

Firstly, we extend an $[n, k]$ LCD code to an $[n+1, k]$ LCD code.

\begin{thm}\label{evenexe}
    Let $C$ be a binary $[n,k]$ $\LCDeo$ code with a generator matrix $G$.
For any $\bm{y} \in \mathbb{F}_2^k$, the matrix $G'=(G \ \bm{y}^T)$ is a generator matrix of an $[n+1,k]$ $\LCD$ code.
\begin{proof}
Without loss of generality, let the rows $\{\bm{b}_1,\bm{b}_1',...,\bm{b}_{k/2},\bm{b}_{k/2}'\}$ of $G$ be a symplectic basis of $C$.
Obviously, replacing $\bm{b}_i$ or $\bm{b}_i'$ by $\bm{b}_i + \bm{b}_i'$ in $\{\bm{b}_1,\bm{b}_1',...,\bm{b}_{k/2},\bm{b}_{k/2}'\}$, for any $1 \le i \le \frac{k}{2}$,
then the transformed vectors is also a symplectic basis of $C$.

For any $\bm{y}=(y_1,y_1',...,y_{k/2},y_{k/2}') \in \mathbb{F}_2^k$,
we do elementary row transform action on $(G \ \bm{y}^T)$ by the following two steps:

\textbf{Step 1}:
If $(y_i,y_i')=(1,0)$, we replace $(\bm{b}_i',y_i')$ by $(\bm{b}_i', y_i') + (\bm{b}_i,y_i)$.

If $(y_i,y_i')=(0,1)$, we replace $(\bm{b}_i,y_i)$ by $(\bm{b}_i', y_i') + (\bm{b}_i,y_i)$.

Denote the transformed matrix by $(G_1 \ \bm{y}'^T)$, we know $G_1G_1^T=\diag(J_2,...,J_2)_{\frac{k}{2}}$ and $y_i=y_i'$ in $\bm{y}'$ for $1 \le i \le k/2$.

\textbf{Step 2}:
We transform $(G_1 \ \bm{y}'^T)$ to $(G_2 \ \bm{y}''^T)$ such that $\bm{y}'$ to \begin{tikzpicture}[baseline=(current bounding box.base)]
    \draw(0,0.12)node{$\bm{y}''=(1,...,1,0,...,0)$};
    \draw(-0.18,-0.16)node{$\underbrace{\rule{10mm}{0mm}}$}(-0.14,-0.46)node{$h$};
\end{tikzpicture}, where $h=wt(\bm{y}')$ and $G_2G_2^T=\diag(J_2,...,J_2)_{\frac{k}{2}}$.
 We have $(G_2 \ \bm{y}''^T)(G_2 \ \bm{y}''^T)^T=
 \begin{pmatrix}
     A &  \\
      &  B
 \end{pmatrix}$,
 where $A=
 \begin{pmatrix}
     I_2 & E_2 &    \cdots &  E_2 \\
     E_2 & I_2 &    \cdots & E_2 \\
     \vdots & \vdots  & \ddots &  \vdots \\
     E_2 & E_2 &  \cdots  & I_2
 \end{pmatrix}$ and $B=\diag(J_2,...,J_2)$.
Note that the matrices $A$ and $B$ are both nonsingular, so $(G \ \bm{y}^T)$ is a generator matrix of an $[n+1, k]$ $\LCD$ code by Lemma~\ref{malcd}.
\end{proof}
\end{thm}

\begin{lem}\label{oddexe}
    Let $C$ be a binary odd-like $[n,k]$ $\LCD$ code with a generator matrix
        \renewcommand\arraystretch{0.8}
        $G=\begin{pmatrix}
        \bm{g}_1  \\
        \vdots \\
        \bm{g}_k  \end{pmatrix}$, where $\langle \bm{g}_k, \bm{g}_i \rangle=0$ for $i=1,...,k-1$.
For any $(y_1,...,y_{k-1}) \in \mathbb{F}_2^{k-1}$, there exists $y_k \in \mathbb{F}_2$ such that $G'=(G \ \bm{y}^T)$ is a generator matrix of an $[n+1,k]$ $\LCD$ code,
where $\bm{y}=(y_1,...,y_{k-1},y_k)$.
\begin{proof}
    Since $\langle \bm{g}_k, \bm{g}_i \rangle=0$ for $i=1,...,k-1$ and $C$ is LCD, we have $D = \langle \bm{g}_1,...,\bm{g_{k-1}} \rangle_{\mathbb{F}_2}$ is a subcode of $C$ and LCD code, and $wt(\bm{g}_k)$ is odd.

If $D$ is even-like, then for any $(y_1,...,y_{k-1}) \in \mathbb{F}_2^{k-1}$, $\{(\bm{g}_1,y_1),...,(\bm{g_{k-1}},y_{k-1})\}$ is a basis of an $[n+1,k-1]$ LCD code by Theorem~\ref{evenexe}.
And then \\$\{(\bm{g}_1,y_1),...,(\bm{g_{k-1}},y_{k-1}),(\bm{g}_k,0)\}$ forms a basis of an $[n+1,k]$ LCD code since $\langle (\bm{g}_k,0) , (\bm{g}_i,y_i)\rangle=0$ for $i=1,...,k-1$ and $wt((\bm{g}_k,0))$ is odd.

If $D$ is odd-like,
without loss of generality, let $\bm{g}_1,...,\bm{g_{k-1}},\bm{g}_k$ be pairwise orthogonal.
For any $(y_1,...,y_{k-1}) \in \mathbb{F}_2^{k-1}$, we can choose $y_n \in \mathbb{F}_2$
such that $wt(\bm{y})$ is even.
Permuting the rows of $(G \ \bm{y}^T)$, we denote the transformed matrix by $(G_1 \ \bm{y}'^T)$ such that
\begin{tikzpicture}[baseline=(current bounding box.base)]
    \draw(0,0.12)node{$\bm{y}'=(1,...,1,0,...,0)$};
    \draw(-0.24,-0.16)node{$\underbrace{\rule{10mm}{0mm}}$}(-0.2,-0.46)node{$h$};
\end{tikzpicture}, where $h=wt(\bm{y})$. We have
\vspace{-0.4cm}
\setlength{\arraycolsep}{1pt}
\renewcommand\arraystretch{0.8}
\[
    (G_1 \ \bm{y}'^T)(G_1 \ \bm{y}'^T)^T=
    \begin{pmatrix}
        J_h-I_h &  \\
         &  I_{k-h}
    \end{pmatrix}.
\]
Note that $(G_1 \ \bm{y}'^T)(G_1 \ \bm{y}'^T)^T$ is nonsingular since $h$ is even. Therefore, $(G \ \bm{y}^T)$ is a generator matrix of an $[n+1,k]$ LCD code.
\end{proof}
\end{lem}

\begin{thm}\label{leoon}
    Let $C$ be a binary odd-like $[n,k]$ $\LCD$ code with a generator matrix $G$ and $GG^T=I_k$.
Then for any $\bm{y}=(y_1,...,y_k) \in \mathbb{F}_2^k$, the matrix $G'=(G \ \bm{y}^T)$ is a generator matrix of an $[n+1,k]$ $\LCD$ code if and only if $wt(\bm{y})$ is even.
\begin{proof}
    Permuting the rows of $(G \ \bm{y}^T)$, we denote the transformed matrix by $(G_1 \ \bm{y}'^T)$ such that
\begin{tikzpicture}[baseline=(current bounding box.base)]
    \draw(0,0.12)node{$\bm{y}'=(1,...,1,0,...,0)$};
    \draw(-0.24,-0.16)node{$\underbrace{\rule{10mm}{0mm}}$}(-0.2,-0.46)node{$h$};
\end{tikzpicture}, where $h=wt(\bm{y})$. We have
\vspace{-0.4cm}
\setlength{\arraycolsep}{1pt}
\renewcommand\arraystretch{0.8}
\[
    (G_1 \ \bm{y}'^T)(G_1 \ \bm{y}'^T)^T=
    \begin{pmatrix}
        J_h-I_h &  \\
         &  I_{k-h}
    \end{pmatrix}.
\]
Therefore $(G \ \bm{y}^T)$ is a generator matrix of an $[n+1,k]$ LCD code if and only if $(G_1 \ \bm{y}'^T)(G_1 \ \bm{y}'^T)^T$ is nonsingular if and only if $h$ is even.
\end{proof}
\end{thm}

In the following, we extend the generator matrix of an LCD code by one column and one row.

\begin{lem}\label{lemfr}
    Let $C$ be a binary $[n,k]$ linear code with a generator matrix $G$ and a parity-check matrix $H$.
    If the first $k$ columns of $G$ are linearly independent, then the last $n - k$ columns of $H$ are linearly independent.
\begin{proof}
    We prove this lemma by contradiction. Let $\{\bm{l}_1,...,\bm{l}_n \}$ be the column vectors of $H$, if the last $n - k$ columns of $H$ are not linearly independent,
then there exists $\bm{0}_{n-k} \neq (x_{k+1},...,x_n) \in \mathbb{F}_2^{n-k}$, such that $x_{k+1}\bm{l}_{k+1}+\cdots+x_n\bm{l}_{n}=\bm{0}_{n-k}^T$.
This means that $(0,...,0,x_{k+1},...,x_n) \in C$, the codewords of $C$ with first $k$ coordinates all being zeros are only $\bm{0}_n$, since the first $k$ columns of $G$ are linearly independent.
This contradicts with $(x_{k+1},...,x_n) \neq \bm{0}_{n-k}$.
\end{proof}
\end{lem}

\begin{thm}\label{simpext}
    Let $C$ be a binary $[n,k]$ $\LCD$ code with a generator matrix $G$ and a parity-check matrix $H$,
and let the first $k$ columns of $G$ be linearly independent.
Let $\bm{x} = (x_{k+1},...,x_n) \in \mathbb{F}_2^{n-k}, \bar{\bm{x}}=(0,...,0,\bm{x}) \in \mathbb{F}_2^n$,
$C(\bm{x})$ be a binary linear code with a generator matrix
\begin{equation}\label{equ3}
    G(\bm{x}) = \begin{pmatrix}
        \bar{\bm{x}} & 1 \\
        G  &0
    \end{pmatrix}.
\end{equation}

\noindent (1) If $C$ is an $\LCDoe$ code, then for any $\bm{x} = (x_{k+1},...,x_n) \in \mathbb{F}_2^{n-k}$, $C(\bm{x})$ is an $\LCD$ code.

\noindent (2) Otherwise, $C(\bm{x})$ is an $\LCD$ code if and only if $\bm{1}_{n-k}\bar{H}\bm{x}^T=0$.(where $\bar{H}$ is the matrix formed by the last $n-k$ columns of $H$ and $HH^T = I_{n-k}$)
\begin{proof}
    Let $A=\begin{pmatrix}
        \bm{r}_1 \\
        \vdots \\
        \bm{r}_{n-k}
    \end{pmatrix}$ be a matrix formed by the last $n - k$ columns of $H$.
    By Lemma \ref{lemfr}, we know the matrix $A$ is invertible.

(1) By Theorem~\ref{evenexe}, for any $\bm{y}=(y_1,...,y_{n-k})$, $(H \ \bm{y}^T)$ is a generator matrix of an $[n+1,n-k]$ LCD code.
Then its dual code has a generator matrix form as shown in (\ref{equ3}),
and it satisfies $\langle \bm{r}_t, \bm{x} \rangle =y_t, t=1,...,n-k$, that is $A\bm{x}^T=\bm{y}^T$,
and then $\bm{x}^T=A^{-1}\bm{y}^T$, the result then follows from the arbitrariness of $\bm{y}$.

(2) We select a parity-check matrix $H$ of $C$ such that $HH^T = I_{n-k}$.
By Theorem~\ref{leoon}, $\bm{y}=(y_1,...,y_{n-k}) \in \mathbb{F}_2^{n-k}$,
$(H\ \bm{y}^T)$ is a generator matrix of an LCD code if and only if $\langle \bm{y}, \bm{1}_{n-k} \rangle=0$.
Then its dual code has a generator matrix form  as in (\ref{equ3}), and it satisfies $\langle \bm{r}_t, \bm{x} \rangle = y_t, t=1,...,n-k$,
that is $A\bm{x}^T = \bm{y}^T$ and then $\bm{1}_{n-k}A\bm{x}^T = 0$.
\end{proof}
\end{thm}

We can also extend linear codes with dimension of the hull $\ge 1$ to obtain LCD codes.

\begin{thm}\label{thseff}
    Let $C$ be an $[n, k]$ linear code with generator matrix $G$ and $\dim(\Hull(C))=s \ge 1$. Let
$\bm{x} \in \mathbb{F}_2^n$ and $C(\bm{x})$ be a linear code with generator matrix as in (\ref{equ2}).
If there exists $\bm{x}' \in \Hull(C)$ such that $\langle \bm{x}, \bm{x}'\rangle=1$, then $C(\bm{x})$ is an $[n+1, k+1]$ code with $\dim(\Hull(C(\bm{x})))=s-1$.
In particular, if $\dim(\Hull(C))=1$, then $C(\bm{x})$ is an $[n+1, k+1]$ $\LCD$ code.
\begin{proof}
    By Lemma~\ref{carleoo}, let $\{\bm{g}_1',...,\bm{g}_s',...,\bm{g}_k'\}$ be a basis of $C$ which consists a generator matrix $G'$ and $\{\bm{g}_1',...,\bm{g}_s'\}$ be a basis of $\Hull(C)$ such that $G'G'^T$ has the form as
        \setlength{\arraycolsep}{1pt}
        \renewcommand\arraystretch{0.8}
        \[
            G'G'^T=
            \begin{pmatrix}
                0_s & 0 \\
                0 & A
            \end{pmatrix}, \ \text{where}\
            A=\begin{cases}
                \diag(J_2,...,J_2)_{\frac{k-s}{2}},  & \text{if $C$ is even-like}, \\
                I_{k-s},  & \text{if $C$ is odd-like}.
            \end{cases}
        \]
Without loss of generality, we can suppose $\langle \bm{g}_1', \bm{x} \rangle=1$. Let $\bm{g}_1=\bm{g}_1'$, $\bm{g}_i = \bm{g}_i' + \langle \bm{g}_i', \bm{x} \rangle \bm{g}_1'$ for $2 \le i \le k$,
we have $\langle \bm{g}_1,\bm{x} \rangle=1$, $\langle \bm{g}_i, \bm{x} \rangle=0$ for $2 \le i \le k$ and $\langle \bm{g}_u,\bm{g}_v \rangle = \langle \bm{g}_u',\bm{g}_v' \rangle$ for $1 \le u,v \le k$.
Then $\{(1,\bm{x}),(0,\bm{g}_1),...,(0,\bm{g_s}),...,(0,\bm{g}_k)\}$ is a basis of $C(\bm{x})$ which consists a generator matrix $G(\bm{x})$ and then
    \setlength{\arraycolsep}{1pt}
    \renewcommand\arraystretch{0.8}
    \[
        G(\bm{x})G(\bm{x})^T=
        \begin{pmatrix}
            \delta & 1 & 0 & 0 \\
            1 & 0 & 0 & 0 \\
            0 & 0 & 0_{s-1} & 0 \\
            0 & 0 & 0 & A
        \end{pmatrix}, \ \text{where}\
        A=\begin{cases}
            \diag(J_2,...,J_2)_{\frac{k-s}{2}},  & \text{if $C$ is even-like}, \\
            I_{k-s},  & \text{if $C$ is odd-like},
        \end{cases}
    \]
    $\delta=\begin{cases}
        0,  & \text{if $wt(\bm{x})$ is odd}, \\
        1,  & \text{if $wt(\bm{x})$ is even}. \end{cases}$

    We have $\dim(\Hull(C(\bm{x})))=(k+1)-rank(G(\bm{x})G(\bm{x})^T)=s-1.$
\end{proof}
\end{thm}

\begin{coro}\label{hhbit}
    Let $C$ be an $[n, k]$ linear code with $\dim(\Hull(C))=s \ge 1$ and $G$ be a generator matrix of $C$. Let
$\bm{x} \in \mathbb{F}_2^n$, $\bm{y}=(\langle \bm{x},\bm{r}_1 \rangle,...,\langle \bm{x},\bm{r}_k \rangle)$ where $\bm{r}_i$ is the $i$th row of $G$ and let $C(\bm{y})$ be a linear code with generator matrix $(G \ \bm{y}^T)$.
If there exists $\bm{x}' \in \Hull(C)$ such that $\langle \bm{x}, \bm{x}'\rangle=1$, then $C(\bm{y})$ is an $[n+1, k]$ code with $\dim(\Hull(C))=s-1$.
In particular, if $\dim(\Hull(C))=1$, then $C(\bm{y})$ is an $[n+1, k]$ $\LCD$ code.
\end{coro}
\begin{proof}
    The dual code $C^{\bot}$ of $C$ and the vector $\bm{x}$ satisfy the conditions of Theorem~\ref{thseff},
$C^{\bot}$ can be extended to an $[n+1,n-k+1]$ linear code with $s-1$ dimension hull.
So the code $C$ can be extended to $C(\bm{y})$ with parameters $[n+1, k]$ and $s-1$ dimension hull.
\end{proof}

By this way, we can extend code length for any linear code to obtain an LCD code.
\begin{coro}
    Let $C$ be an $[n, k, d]$ linear code with $\dim(\Hull(C))=s$, then there exists an $[n+s,k,\ge d]$ $\LCD$ code.
\end{coro}

Compared to using Lemma~\ref{pross3}(2), by using the following theorem, we can obtain more distinct LCD codes with parameters $[n+2,k,\ge d+1]$.
 We provide an example to illustrate this point at the end of this section.

\begin{thm}\label{muccs}
    If $C$ is an $[n, k, d]$ $\LCD$ code, where $k$ and $d$ are odd,
then we can construct at most $2^{k-1}$ different $[n+2,k,\ge d+1]$ $\LCD$ codes.
\begin{proof}
    We extend the code $C$ to obtain an $[n+1,k,d+1]$ even-like code $C'$ with $\dim(\Hull(C'))=1$.
    By Lemma~\ref{carleoo}, we find a generator matrix
    \renewcommand\arraystretch{0.8}
    $G'=\begin{pmatrix}
    \bm{g}_1'  \\
    \vdots \\
    \bm{g}_k'  \end{pmatrix}$ of $C'$ such that \setlength{\arraycolsep}{1pt}
    \renewcommand\arraystretch{0.8}
    $
        G'G'^T=
        \begin{pmatrix}
            0 & 0 \\
            0 & A
        \end{pmatrix}, \ \text{where}\
        A=\diag(J_2,...,J_2)_{\frac{k-1}{2}}
    $. For any $\bm{y}=(y_1,...,y_k) \in \mathbb{F}_2^k$ with $y_1 = 1$,
    let
    \renewcommand\arraystretch{0.8}
    $(G_1' \ \bm{y}'^T)=\begin{pmatrix}
    (\bm{g}_1', 1 )   \\
    (\bm{g}_2', y_2 ) + (\bm{g}_1', 1 )y_2 \\
    \vdots \\
    (\bm{g}_k', y_k )+ (\bm{g}_1', 1 )y_k  \end{pmatrix}$.
We have \setlength{\arraycolsep}{1pt}
\renewcommand\arraystretch{0.8}
$
    (G_1' \ \bm{y}'^T)(G_1' \ \bm{y}'^T)^T=
    \begin{pmatrix}
        1 & 0 \\
        0 & A
    \end{pmatrix}, \ \text{where}\
    A=\diag(J_2,...,J_2)_{\frac{k-1}{2}}
$, then $(G_1' \ \bm{y}'^T)$ is a generator matrix of an $[n+2,k,\ge d+1]$ LCD code.
We know that $(y_2,...,y_k) \in \mathbb{F}_2^{k-2}$ has $2^{k-1}$ different values corresponding to at most $2^{k-1}$ different LCD codes.
\end{proof}
\end{thm}

\begin{exam}
    Let $C$ be a binary $\LCD$ code with parameters $[43,7,19]$, and with a generator matrix
    \[
        G=\begin{pmatrix}
            1000000110001100101101001101001110001110010 \\
            0100000001101100010010111101110010001101010 \\
            0010010111100100101001010111110000101001101 \\
            0001010001111000010101011111000011011000101 \\
            0000110010110001110011001100010111100111100 \\
            0000001101001101110011001011101000010011100 \\
            0000000000000011111111111111111111111000000 \\
        \end{pmatrix}.
    \]
    With the assistance of computer calculations,
    we construct $64$ different binary $\LCD$ codes with parameters of $[45,7,20]$ from code $C$ in the same way as Theorem~\ref{muccs},
    and these codes are not equivalent to the each other.
\end{exam}

\section{Construction of LCD codes from $\LCDoe$ codes}
In this section, all codes are assumed to be binary.

We can obtain an $[n-1,k,d-1]$ LCD code by puncturing a coordinate of an $[n,k,d](d \ge 2)$ $\LCDeo$ or $\LCDoo$ code.
By Lemmas~\ref{lema2} and \ref{pro34}(1), the punctured code of an $[n, k, d] (d \ge 2)$ $\LCDoe$ code by puncturing one coordinate position can not be LCD.

In this section, by puncturing two coordinate positions of an $\LCDoe$ code and then extend it by one column,
we obtain a sufficient condition for constructing $[n-1,k,d-1]$ LCD codes from an $[n, k, d] (d \ge 3)$ $\LCDoe$ code.

\begin{lem}\label{lemPPON}
    Let $C$ be an $[n,k]$ code with $\dim(\Hull(C))=1$. Suppose $\bm{0}_n \neq \bm{g}_1 \in \Hull(C)$,
and $\{v_1,...,v_m\}$ are the all nonzero coordinates of $\bm{g}_1$.
Then the punctured code $C^v$ of $C$ on any one coordinate $v \in \{v_1,...,v_m\}$ is $\LCD$.
\begin{proof}
    There exist $\bm{g}_2,...,\bm{g}_k$, such that ${\bm{g}_1,...,\bm{g}_k}$ form a basis of $C$. Let \renewcommand\arraystretch{0.8}
    $G=\begin{pmatrix}
    \bm{g}_1  \\
    \vdots \\
    \bm{g}_k  \end{pmatrix}$, then $GG^T=\begin{pmatrix}
        0 &  \\
         &  A
    \end{pmatrix}$, where $A$ is an invertible symmetric matrix. Without loss of generality, we can suppose the first coordinate of $\bm{g}_1$ is not $0$.
    Let $$\bm{g}_1' = \bm{g}_1, \bm{g}_i'= \begin{cases}
        \bm{g}_i+\bm{g}_1, & \text{if the first coordinate of $\bm{g}_i$ equal to $1$}, \\
        \bm{g}_i, & \text{otherwise},
    \end{cases}(i=2,...,k)$$ and let $G'=\begin{pmatrix}
        \bm{g}_1' \\
        \vdots \\
        \bm{g}_k'  \end{pmatrix}$. We have $G'G'^T=GG^T=\begin{pmatrix}
        0 &  \\
         &  A
    \end{pmatrix}$, where $A$ is an invertible symmetric matrix.
    Denote $C^{1}$ as the punctured code of $C$ on first coordinate, $G'^{1}$ as the matrix obtained by deleting the first column of $G'$.
    We know $G'^{1}$ is a generator matrix of $C^{1}$, and we have $G'^{1}(G'^{1})^T=\begin{pmatrix}
        1 &  \\
         &  A
    \end{pmatrix}$, which is invertible, thus $C^1$ is LCD.
\end{proof}
\end{lem}

\begin{lem}\label{ogk}
    Let $C$ be an $[n+1,k,d](d \ge 3)$ $\LCDoe$ code. For any coordinate $1 \le {u} \le n+1$ there exist at least $d-1$ coordinates $\{v_1,...,v_m\}$ and $u \notin \{v_1,...,v_m\}$,
such that the punctured code $C^{u,v}$ is $\LCD$ for any $v \in \{v_1,...,v_m\}$.
\begin{proof}
    Without loss of generality, we may assume $u=1$.
    Let $G=(g_{ij})$ be a generator matrix of $C$ with $GG^T=I_k$. Let $\bm{g}_i$ be the $i$th row of $G$ and $\bm{l}_j$ be the $j$th column of $G$,
    we can suppose \begin{tikzpicture}[baseline=(current bounding box.base)]
        \draw(0,0.12)node{$\bm{l}_1^T=(1,...,1,0,...,0)$};
        \draw(-0.18,-0.16)node{$\underbrace{\rule{10mm}{0mm}}$}(-0.14,-0.46)node{$h$};
    \end{tikzpicture}.
    By Lemma~\ref{allonev}, we know that $h$ is odd.
    Let $C^1$ denote the punctured code of $C$ on the first coordinate and
    let $G^1$ be the matrix by deleting the first column of $G$.
    Then we have $rank(E_h-I_h)=h-1$ and $G^1(G^1)^T=\diag(E_h-I_h,I_{n+1-h})$ is singular. By \cite[Proposition 3.1]{KS2018}, $\dim(\Hull(C^1))=k-rank(G^1(G^1)^T)=1$.
    It is easy to see that $\bm{g}_1^1+\cdots+\bm{g}_h^1 \in \Hull(C^1)$ and $wt(\bm{g}_1^1+\cdots+\bm{g}_h^1) \ge d-1$,
    where $\bm{g}_t^1$ is the vector which deletes the first coordinate of $\bm{g}_t$ for $t=1,...,h$.
    Suppose $\{1,v_1,...,v_m\}$ are the all nonzero coordinates of the vector $\bm{g}_1+\cdots+\bm{g}_h$.
    By Lemma~\ref{lemPPON}, we have the punctured code $C^{1,v}$ is LCD for any $v \in \{v_1,...,v_m\}$.
\end{proof}
\end{lem}

We can easily draw the following conclusion.
\begin{thm}
For any code length $n$ and dimension $k$, we have
\[
    d_{\LCD}(n + 1, k) =
        d_{\LCD}(n, k) \ \text{or}\  d_{\LCD}(n, k) + 1 \ \text{or}\  d_{\LCD}(n, k) + 2.
\]
\begin{proof}
    For an $[n+1,k,d](d \ge 2)$ $\LCDeo$ or $\LCDoo$ code, by Lemmas~\ref{pro33} and \ref{pro34}(2), we can obtain an $[n,k,\ge d-1]$ LCD code.
    For an $[n+1,k,d](d \ge 3)$ $\LCDoe$ code, by Lemma~\ref{ogk}, we can obtain an $[n-1,k,\ge d-2]$ LCD code.
    By extending the punctured code with an all-zero column, we can construct an $[n,k,\ge d-2]$ LCD code.
\end{proof}
\end{thm}

Now, we consider how to construct an $[n,k,d-1]$ LCD code from an $\LCDoe$ code with parameters $[n+1,k,d]$ where $k \ge 2$ and $d \ge 3$.

Let $C$ be a binary $[n+1,k,d]$ $\LCDoe$ code with $d \ge 3$ and $k \ge 2$.
By Lemma~\ref{ogk}, there exist two distinct coordinates $1 \le i,j \le n+1$ such that the punctured code $C^{i,j}$ is an $[n-1,k, \ge d-2]$ LCD code.
We denote the puncture map on coordinates $i,j$ as
\begin{align*}
    \Omega^{i,j} :C & \longrightarrow C^{i,j},\\
    \bm{x}=(x_1,...,x_{n+1}) &\longmapsto \bm{x}^{i,j}=(...,x_{i-1},x_{i+1},...,x_{j-1},x_{j+1},...),
\end{align*}
where $\bm{x}^{i,j}$ is the vector by deleting the coordinates $i,j$ of $\bm{x}$.
The puncture map $\Omega^{i,j}$ is a linear isomorphism, since the number of puncturing coordinates is less than the minimum distance of the code being punctured.
We know $C^{i,j}$ is an $[n-1,k,\ge d-2]$ LCD code.

Since $\bm{1}_{n+1} \in C$, then $\bm{1}_{n-1} \in C^{i,j}$, we have $C^{i,j}$ is $\LCDoe$.
Now, we consider the case of $C^{i,j}$ being an $[n-1,k, d-2]$ $\LCDoe$ code.
Let
\[
    C_d=\langle \bm{a} \in C : wt(\bm{a})=d \rangle_{\mathbb{F}_2},
    C^{i,j}_{d-2}=\langle \bm{c}^{i,j} \in C^{i,j} : wt(\bm{c}^{i,j})=d-2 \rangle_{\mathbb{F}_2}.
\]
Suppose $\{\bm{a}_1^{i,j},...,\bm{a}_r^{i,j}\}$ are all the minimum weight codewords of $C^{i,j}$, then $ C^{i,j}_{d-2} = \langle \bm{a}_1^{i,j},...,\bm{a}_r^{i,j}\rangle_{\mathbb{F}_2}$.

\begin{lem}\label{lemm11}
    Assume the notations are given above. If there exists an odd weight codeword $\bm{c}^{i,j} \in (C^{i,j} \backslash C^{i,j}_{d-2}) \cap {C^{i,j}_{d-2}}^{\bot}$,
then $C^{i,j}$ can be extended to an $[n,k,d-1]$ $\LCD$ code.
\begin{proof}
Note that $wt(\bm{c}^{i,j})$ is odd, by Lemma~\ref{impp},
    $\bm{c}^{i,j}$ can be extended to a basis $\{\bm{c}_1^{i,j},...,\bm{c}_{k-1}^{i,j},\bm{c}^{i,j}\}$ of $C^{i,j}$ such that $\langle \bm{c}^{i,j}, \bm{c}_w^{i,j} \rangle=0$ for $w=1,...,k-1$.
We have $C^{i,j}_{d-2} \subseteq \langle \bm{c}_1^{i,j},...,\bm{c}_{k-1}^{i,j} \rangle_{\mathbb{F}_2}$ by $\bm{c}^{i,j} \in (C^{i,j} \backslash C^{i,j}_{d-2}) \cap {C^{i,j}_{d-2}}^{\bot}$.

We claim that $\langle (\bm{a}_1^{i,j},1),...,(\bm{a}_r^{i,j},1) \rangle_{\mathbb{F}_2}$ is a linear code with minimum weight $d-1$,
and there exists a vector $(y_1,...,y_{k-1}) \in \mathbb{F}_2^{k-1}$ such that
\[
    \langle (\bm{a}_1^{i,j},1),...,(\bm{a}_r^{i,j},1)\rangle_{\mathbb{F}_2} \subseteq \langle (\bm{c}_1^{i,j},y_1),...,(\bm{c}_{k-1}^{i,j},y_{k-1}) \rangle_{\mathbb{F}_2}.
\]
For the inverse map $(\Omega^{i,j})^{-1}$ of the puncture map $\Omega^{i,j}$, we have
\[
    (\Omega^{i,j})^{-1}(\bm{a}^{i,j})=(...,x_{i-1},1,x_{i+1},...,x_{j-1},1,x_{j+1},...) \text{ and }  wt((\Omega^{i,j})^{-1}(\bm{a}^{i,j}))=d,
\]
for any $\bm{a}^{i,j}=(...,x_{i-1},x_{i+1},...,x_{j-1},x_{j+1},...) \in \{\bm{a}_1^{i,j},...,\bm{a}_r^{i,j}\}$. And
\begin{align*}
    (\Omega^{i,j})^{-1}:\langle \bm{a}_1^{i,j},...,\bm{a}_r^{i,j}\rangle_{\mathbb{F}_2} & \longrightarrow \langle \bm{a_1},...,\bm{a_r} \rangle_{\mathbb{F}_2} \\
    \bm{a^{i,j}} & \longmapsto (\Omega^{i,j})^{-1}(\bm{a^{i,j}}).
\end{align*} is a linear isomorphism. The code $\langle \bm{a_1},...,\bm{a_r}^{i,j} \rangle_{\mathbb{F}_2}$ is a linear subcode of $C_d$.
Similarly
\begin{align*}
    \Upsilon :\langle \bm{a}_1^{i,j},...,\bm{a}_r^{i,j}\rangle_{\mathbb{F}_2} & \longrightarrow \langle (\bm{a}_1^{i,j},1),...,(\bm{a}_r^{i,j},1) \rangle_{\mathbb{F}_2} \\
    \bm{a^{i,j}_t} & \longmapsto (\bm{a^{i,j}_t},1) \text{ for } t=1,...,r.
\end{align*} is a linear isomorphism.
It's easy to see that $\langle (\bm{a}_1^{i,j},1),...,(\bm{a}_r^{i,j},1) \rangle_{\mathbb{F}_2}$ is equivalent to
$\langle \Omega^i(\bm{a_1}),...,\Omega^i(\bm{a_r}) \rangle_{\mathbb{F}_2}$, where $\Omega^i$ is the puncture map on coordinate $i$.
We have $\langle (\bm{a}_1^{i,j},1),...,(\bm{a}_r^{i,j},1) \rangle_{\mathbb{F}_2}$ is a linear code with minimum weight $d-1$.
According to the fundamental results of linear algebra, there exists $(y_1,...,y_{k-1}) \in \mathbb{F}_2^{k-1}$ such that
\[
    \langle (\bm{a}_1^{i,j},1),...,(\bm{a}_r^{i,j},1)\rangle_{\mathbb{F}_2} \subseteq \langle (\bm{c}_1^{i,j},y_1),...,(\bm{c}_{k-1}^{i,j},y_{k-1}) \rangle_{\mathbb{F}_2}.
\]

We claim $\langle (\bm{c}_1^{i,j},y_1),...,(\bm{c}_{k-1}^{i,j},y_{k-1}), (\bm{c}^{i,j},y_k) \rangle_{\mathbb{F}_2}$ is an $[n,k,d-1]$ code, for any $y_k \in \mathbb{F}_2$.
We prove this claim by contradiction. For any fixed $y_k \in \mathbb{F}_2$, if there exists $(\bar{\bm{c}}^{i,j}, \bar{y}) \in \langle (\bm{c}_1^{i,j},y_1),...,(\bm{c}_{k-1}^{i,j},y_{k-1}), (\bm{c}^{i,j},y_k) \rangle_{\mathbb{F}_2}$
with $wt((\bar{\bm{c}}^{i,j}, \bar{y})) \le d-2$, then $wt(\bar{\bm{c}}^{i,j}) \le d-2$ and $\bar{\bm{c}}^{i,j}$ is equal to one vector in $\{\bm{a}_1^{i,j},...,\bm{a}_r^{i,j}\}$.
We have $\bar{y}=1$ and $wt((\bar{\bm{c}}^{i,j}, \bar{y}))=d-1$, contradicts with $wt((\bar{\bm{c}}^{i,j}, \bar{y})) \le d-2$.

By Lemma~\ref{oddexe}, we can choose $y_k \in \mathbb{F}_2$ appropriately, such that \\
$\langle (\bm{c}_1^{i,j},y_1),...,(\bm{c}_{k-1}^{i,j},y_{k-1}), (\bm{c}^{i,j},y_k) \rangle_{\mathbb{F}_2}$ is LCD.
Therefore, $C^{i,j}$ can be extended to an $[n,k,d-1]$ LCD code.
\end{proof}
\end{lem}

Below, we will provide further characterization by studying subcodes of an $\LCDoe$ code.

\begin{lem}\label{impp2}
    Let $C$ be a binary odd-like $[n,k]$ $\LCD$ code, $D$ be an $[n, k_1 < k]$ subcode of $C$, $s=\dim(\Hull(D))$, and $t=k-k_1-s$.
Then there exists a generator matrix
    \setlength{\arraycolsep}{1pt}
    \renewcommand\arraystretch{0.8}
    $G=\begin{pmatrix}
    \bm{g}_1  \\
    \vdots \\
    \bm{g}_k  \end{pmatrix}$
    of $C$, such that $\{\bm{g}_1,...,\bm{g}_{k_1}\}$ is a basis of $D$ and $GG^T$ has the following form
\setlength{\arraycolsep}{1pt}
\renewcommand\arraystretch{0.8}
\begin{equation}\label{eqq1}
    GG^T=
    \begin{pmatrix}
        0_s & 0 & I_s & 0 \\
        0 & A_1 & 0  &  0 \\
        I_s & 0 &  A_2 &  0 \\
        0 & 0  & 0 &  A_3
    \end{pmatrix}, \ \text{where}\
    A_1=\begin{cases}
        \diag(J_2,...,J_2)_{\frac{k_1-s}{2}},  & \text{if $D$ is even-like}, \\
        I_{k_1-s},  & \text{if $D$ is odd-like},
    \end{cases}
\end{equation}
\begin{tikzpicture}[baseline=(current bounding box.base)]
    \draw(0,0.12)node{$A_2=\diag(1,...,1,0,...,0)$,};
    \draw(0.18,-0.16)node{$\underbrace{\rule{10mm}{0mm}}$}(0.24,-0.46)node{$s_1$};
\end{tikzpicture}
$   0 \le s_1 \le s,
    A_3=\diag(J_2,...,J_2)_{\frac{t}{2}} \ \text{or} \ I_t.
$
\begin{proof}
    By Lemma~\ref{carleoo}, there exists a generator matrix
    \setlength{\arraycolsep}{1pt}
    \renewcommand\arraystretch{0.8}
    $G_1=\begin{pmatrix}
    \bm{g}_1  \\
    \vdots \\
    \bm{g}_{k_1}  \end{pmatrix}$
    of $D$, such that
    \setlength{\arraycolsep}{1pt}
    \renewcommand\arraystretch{0.8}
    \[G_1G_1^T=\begin{pmatrix}
    0_s & 0 \\
    0 & A_1 \end{pmatrix},\text{where }
    A_1=\begin{cases}
        \diag(J_2,...,J_2)_{\frac{k_1-s}{2}},  & \text{if $D$ is even-like,} \\
        I_{k_1-s},  & \text{if $D$ is odd-like}.
    \end{cases}\]
    And there exist $\bm{g}_{k_1+1}',...,\bm{g}_k' \in C$ such that
    $G'=\begin{pmatrix}
    \bm{g}_1  \\
    \vdots \\
    \bm{g}_{k_1} \\
    \bm{g}_{k_1+1}' \\
    \vdots \\
    \bm{g}_{k}' \end{pmatrix}$ is a generator matrix of $C$.
    We write $G'G'^T$ to be a partitioned matrix as $L_0$, where $N_{10}$ is an $s \times s$ matrix, the sizes of other blocks of $L_0$ can also be determined.
    We get matrices $L_1,L_2,L_3,L_4$ by using congruent transformation on $L_0=G'G'^T$.
    \begin{center}
        $L_0=\begin{pmatrix}
            0_s & 0 & N_{10}^T & N_{20}^T \\
            0 & A_1 & N_{30}^T & N_{40}^T \\
            N_{10} & N_{30} & N_{50} & N_{60}^T \\
            N_{20} & N_{40} & N_{60} & N_{70}
        \end{pmatrix}$,
        $L_1=\begin{pmatrix}
            0_s & 0 & N_{10}^T & N_{20}^T \\
            0 & A_1 & 0 & 0 \\
            N_{10} & 0 & N_{51} & N_{61}^T \\
            N_{20} & 0 & N_{61} & N_{71}
        \end{pmatrix}$,
        $L_2=\begin{pmatrix}
            0_s & 0 & I_s & 0 \\
            0 & A_1 & 0 & 0 \\
            I_s & 0 & N_{52} & N_{62}^T \\
            0 & 0 & N_{62} & N_{72}
        \end{pmatrix}$,
        $L_3=\begin{pmatrix}
            0_s & 0 & I_s & 0 \\
            0 & A_1 & 0 & 0 \\
            I_s & 0 & N_{53} & 0 \\
            0 & 0 & 0 & N_{72}
        \end{pmatrix}$,
        $L_4=\begin{pmatrix}
            0_s & 0 & I_s & 0 \\
            0 & A_1 & 0 & 0 \\
            I_s & 0 & A_2 & 0 \\
            0 & 0 & 0 & A_3
        \end{pmatrix}$.
    \end{center}

    Let $P_1=\begin{pmatrix}
        I_s & 0 & 0 & 0 \\
        0 & I_{k_1-s} & 0 & 0 \\
        0 & N_{30}{A_1}^{-1} & I_s & 0 \\
        0 & N_{40}{A_1}^{-1} & 0 & I_t
    \end{pmatrix}$, then $L_1=P_1L_0P_1^T$, where
    \[ \begin{pmatrix}N_{51} & N_{61}^T \\
         N_{61} & N_{71}
    \end{pmatrix}=\begin{pmatrix}N_{30}A_1^{-1}N_{30}^T+N_{50} & N_{30}A_1^{-1}N_{40}^T + N_{60}^T \\
    N_{40}A_1^{-1}N_{30}^T+N_{60} & N_{40}A_1^{-1}N_{40}^T+N_{70}
\end{pmatrix}. \]

    Since $L_1$ is invertible, then
    rank($\begin{pmatrix}
    N_{10} \\
    N_{20} \end{pmatrix})=s$,
    there exists a matrix $H_1 \in GL_{(k-k_1)}(\mathbb{F}_2)$ such that
        $H_1\begin{pmatrix}
        N_{10} \\
        N_{20} \end{pmatrix}=\begin{pmatrix}
            I_s \\
            0 \end{pmatrix}$.
    Let $P_2 = \begin{pmatrix}
        I_{k_1} &  0 \\
        0 & H_1 \\
        \end{pmatrix}$, we have $L_2=P_2L_1P_2^T$, where $\begin{pmatrix}N_{52} & N_{62}^T \\
            N_{62} & N_{72}
        \end{pmatrix}=H_1\begin{pmatrix}N_{51} & N_{61}^T \\
        N_{61} & N_{71}
    \end{pmatrix}H_1^T$.

    Let $P_3 =\begin{pmatrix}
        I_s & 0 & 0 & 0 \\
        0 & I_{k_1-s} & 0 & 0 \\
        N_{52}^* & 0 & I_s & 0 \\
        N_{62} & 0 & 0 & I_t
    \end{pmatrix}$, then $L_3=P_3L_2P_3^T$. The lower triangular position (excluding diagonal positions) elements  of the matrix $N_{52}^*$ are equal to the corresponding position elements of the matrix $N_{52}$,
    and other position elements of $N_{52}^*$ are equal to $0$. The diagonal position elements of the matrix $N_{53}$ are equal to the corresponding position elements of the matrix $N_{52}$,
    and the other position elements of $N_{53}$ are equal to $0$.

    Since $N_{53}$ and $N_{73}$ are symmetric matrices, there exist matrices $M_1 \in GL_s(\mathbb{F}_2)$ and $M_2 \in GL_t(\mathbb{F}_2)$
    such that $M_1N_{53}M_1^T=A_2$ and $M_2N_{72}M_2^T=A_3$. Let $P_4=\diag(I_{k_1},M_1,M_2)$, then $L_4=P_4L_3P_4^T$.

    Let $G = P_4P_3P_2P_1G'$, then we can check that $GG^T$ meets the requirement.

    It is obvious that $t \ge 0$. We are done.
\end{proof}
\end{lem}

\begin{remark}\label{remmm}
    In Lemma~\ref{impp2}, if $A_1=\diag(J_2,...,J_2)_{\frac{k_1-s}{2}}$ and $A_3=\diag(J_2,...,J_2)_{\frac{t}{2}}$, then
we have $s_1 \ge 1$, since $C$ is odd-like, $C$ contains odd weight codewords.
\end{remark}

In Lemma~\ref{impp2}, by observing the matrix $GG^T$ in (\ref{eqq1}), we can divide the vectors in the set $\{\bm{g}_1,...,\bm{g}_k\}$ into three types.

    \textbf{Type 1}. $\langle \bm{g}_w, \bm{g}_w \rangle = 1$, $\bm{g}_w \in \{\bm{g}_1,...,\bm{g}_k\}$,
    $\bm{g}_w$ is orthogonal to vectors of $\{\bm{g}_1,...,\bm{g}_k\} \backslash\{\bm{g}_w\}$.

    \textbf{Type 2}.  $\langle \bm{g}_l, \bm{g}_m \rangle = 1$, $\bm{g}_l, \bm{g}_m \in \{\bm{g}_1,...,\bm{g}_k\}$,
    $\bm{g}_l, \bm{g}_m$ are orthogonal to vectors of $\{\bm{g}_1,...,\bm{g}_k\} \backslash\{\bm{g}_l,\bm{g}_m\}$ and only one of $\bm{g}_l, \bm{g}_m$ is even weight.

    \textbf{Type 3}. $\langle \bm{g}_u, \bm{g}_v \rangle = 1$, $\bm{g}_u, \bm{g}_v \in \{\bm{g}_1,...,\bm{g}_k\}$,
    $\bm{g}_u, \bm{g}_v$ are even weights and they are orthogonal to vectors of $\{\bm{g}_1,...,\bm{g}_k\}\backslash\{\bm{g}_u,\bm{g}_v\}$.

    Assume $C$ is an $\LCDoe$ code, we know $\bm{1}_n \in C$. Now, we consider how to obtain $\bm{1}_n$ from $\{\bm{g}_1,...,\bm{g}_k\}$.

    Let $V_i=\{\bm{v} \in \{\bm{g}_1,...,\bm{g}_k\} : \bm{v} \text{ is Type $i$} \}, i=1,2,3$, let $\{\bm{g}_{k+1},...,\bm{g}_n\}$ be a basis of $C^{\bot}$.

    We have
    $\langle V_1 \cup V_2 \rangle_{\mathbb{F}_2}$ is $\LCDoe$, since $\langle V_3 \cup \{\bm{g}_{k+1},...,\bm{g}_n\} \rangle_{\mathbb{F}_2}$ is even-like and
    $\langle V_1 \cup V_2 \rangle_{\mathbb{F}_2} \cap \langle V_3 \cup \{\bm{g}_{k+1},...,\bm{g}_n\} \rangle_{\mathbb{F}_2} = \bm{0}_n$. We have $\bm{1}_n \in \langle V_1 \cup V_2 \rangle_{\mathbb{F}_2}$.

    If $s_1 = 0$, there does not exist a vector with Type $2$ in $\{\bm{g}_1,...,\bm{g}_k\}$, so $V_1$ forms an orthonormal basis of an $\LCDoe$ code.

    If $s_1 > 0$, let $\bm{g}_1' = \bm{g}_1 + \bm{g}_{k_1+1}, \cdots, \bm{g}_{s_1}' = \bm{g}_{s_1}+\bm{g}_{k_1+s_1}$, \\
    then $V_1 \cup \{\bm{g}_1',...,\bm{g}_{s_1}', \bm{g}_{k_1+1},...,\bm{g}_{k_1+s_1}\}$ forms an orthonormal basis of an $\LCDoe$ code.
    By Lemma~\ref{allonev} we know that for an $[n,k]$ $\LCDoe$ code, $\bm{1}_n$ is equal to the sum of an orthogonal basis of the $\LCDoe$ code.
    We have the following results.

\begin{lem}\label{nott}
    Let $C$ be an $[n,k]$ $\LCDoe$ code, and assume other notations are the same as in Lemma~\ref{impp2}. We have

    \noindent (1) If $D$ is even-like and $A_3=\diag(J_2,...,J_2)_{\frac{t}{2}}$ then $\bm{1}_n = \bm{g}_1+\cdots+\bm{g}_{s_1}$.

    \noindent (2) If $D$ is even-like and $A_3=I_t$ then $\bm{1}_n = \bm{g}_1+\cdots+\bm{g}_{s_1} + \bm{g}_{k_1+s+1}+\cdots+\bm{g}_k$.

    \noindent (3) If $D$ is odd-like and $A_3=\diag(J_2,...,J_2)_{\frac{t}{2}}$ then $\bm{1}_n = \bm{g}_1+\cdots+\bm{g}_{s_1} + \bm{g}_{s+1}+\cdots+\bm{g}_{k_1}$.

    \noindent (4) If $D$ is odd-like and $A_3=I_t$ then $\bm{1}_n = \bm{g}_1+\cdots+\bm{g}_{s_1} + \bm{g}_{s+1}+\cdots+\bm{g}_{k_1}+ \bm{g}_{k_1+s+1}+\cdots+\bm{g}_k$.

\begin{proof}
    We only prove $(1)$. The other statements can be proved similarly.

    In this case, we know there does not exist a vector with Type $1$ among $\{\bm{g}_1,...,\bm{g}_k\}$.
    By Remark~\ref{remmm}, we know that $s_1 \ge 1$. Let $\bm{g}_1' = \bm{g}_1 + \bm{g}_{k_1+1}, \cdots, \bm{g}_{s_1}' = \bm{g}_{s_1}+\bm{g}_{k_1+s_1}$.
    By the above analysis and  Lemma~\ref{allonev}, we have $\bm{1}_n=\bm{g}_1' + \cdots + \bm{g}_{s_1}' + \bm{g}_{k_1+1} + \cdots + \bm{g}_{k_1+s_1}=\bm{g}_1+\cdots+\bm{g}_{s_1}$.
\end{proof}
\end{lem}

\begin{remark}\label{rkkk}
    For Cases (1) and (3) in Lemma~\ref{nott}, we know $\bm{1}_n \in D$. If $A_3$ does not exist we have

    \noindent (i) $\bm{1}_n = \bm{g}_1+\cdots+\bm{g}_{s_1} \in D$ when $D$ is even-like.

    \noindent (ii) $\bm{1}_n = \bm{g}_1+\cdots+\bm{g}_{s_1} + \bm{g}_{s+1}+\cdots+\bm{g}_{k_1} \in D$ when $D$ is odd-like.
\end{remark}

\begin{thm}\label{mmmimmp}
    Assume the notations are the same as in Lemma~\ref{lemm11}. If $\bm{1}_{n-1} \notin C^{i,j}_{d-2}$,
    then there exists an odd weight codeword $\bm{c}^{i,j} \in (C^{i,j} \backslash C^{i,j}_{d-2}) \cap {C^{i,j}_{d-2}}^{\bot}$.
In particular, $C^{i,j}$ can be extended to an $[n,k,d-1]$ $\LCD$ code.
\begin{proof}
    We prove this theorem by contradiction. By Lemma~\ref{impp2}, there exists a basis $\{\bm{g}_1^{i,j},...,\bm{g}_{k_1}^{i,j},...,\bm{g}_k^{i,j}\}$ of $C^{i,j}$ consisting a generator matrix $G$
    where $\{\bm{g}_1^{i,j},...,\bm{g}_{k_1}^{i,j}\}$ is a basis of $C^{i,j}_{d-2}$ such that $GG^T$ has the form as in (\ref{eqq1}).
    Assume there does not exist a vector $\bm{c}^{i,j}$ as above, then $A_3$ does not exist or $A_3=\diag(J_2,...,J_2)_{\frac{t}{2}}$, $t > 0$.
    By Remark~\ref{rkkk}, we know $\bm{1}_{n-1} \in C^{i,j}_{d-2}$, a contradiction.
    Then $C^{i,j}$ can be extended to an $[n,k,d-1]$ LCD code by Lemma~\ref{lemm11}.
\end{proof}
\end{thm}

\begin{thm}\label{immoon}
    Assume the notations are the same as in Lemma~\ref{lemm11}. If $k \not \equiv d \pmod 2$, then there exists an $[n, k, d - 1]$ $\LCD$ code.
\begin{proof}
    If $d$ is even and $k$ is odd, we know $d-2$ is even, $n+1$ is odd, and then $C_{d-2}^{i,j}$
is even-like, $wt(\bm{1}_{n-1})$ is odd and $\bm{1}_{n-1} \notin C_{d-2}^{i,j}$, $C_{d-2}^{i,j}$
can be extended to an $[n, k , d-1]$ LCD code by Theorem~\ref{mmmimmp}.

    If $d$ is odd and $k$ is even,  $C^i$ is the punctured code of $C$ on coordinate $i$ with
parameters $[n ,k ,d - 1]$ and $\bm{1}_n \in C^i$. Let $C^i_{d-1}=\langle\bm{c}^i \in C^i : wt(\bm{c}^i)=d-1\rangle_{\mathbb{F}_2}$.
Since $d- 1$ is even, $n$ is odd and $C_{d-1}^i$ is even, we have $\bm{1}_n \notin C^i_{d-1}$.
And then $\bm{1}_{n-1} \notin C_{d-2}^{i,j}$, by Theorem~\ref{mmmimmp}, $C^{i,j}$ can be extended to an $[n, k, d - 1]$ LCD code.
\end{proof}
\end{thm}

When $k \equiv d \pmod 2$, the condition of Theorem~\ref{mmmimmp} can be satisfied in some cases.
We illustrate it through the following example.

\begin{exam}
    By using the function $BKLC(GF(2), 33, 23)$ of Magma, we obtain a binary $\LCDoe$ code $C$ with parameters $[33,23,5]$.
We puncture the coordinates $\{1,4\}$, and obtain the punctured code $C^{1,4}$, which is a $[31,23,3]$ $\LCD$ code and $\bm{1}_{31} \notin C^{1,4}$.
By Theorem~\ref{mmmimmp}, the code $C^{1,4}$ can be extended to a $[32,23,4]$ $\LCD$ code, which is optimal by the value $d(32,23)=4$, the value can be found in codetables \cite{GTB22}.
\end{exam}

\section{New binary LCD codes}
In this section, we use methods which are given in Sections 2-4 to construct some new binary LCD codes.
All the calculation results of this section can be found in \cite{Wang}. The values of $d_{\LCD}(n,k)$ or the range are presented in Tables 5 and 6,
the corresponding construction methods can be found in Tables 1-4.

By using MAGMA function BKLC, we find binary LCD codes ($41 \le n \le 51$) with parameters
$[41,8,17]$, $[42,8,18]$, $[41,21,9]$, $[41,20,10]$, $[42,14,13]$, $[43,14,14]$, $[43,15,13]$,
$[45,16,14]$, $[49,30,8]$, $[49,37,5]$, $[50,38,5]$, $[51,14,18]$.

Moreover, we use the program Generation of the software package QextNewEdition\cite{BI20}.
This program classifies the linear $[n,k,\ge d]$ codes with dual distance $\ge d^{\bot}$ for given integer $n,k,d$ and $d^{\bot}$.
With the Generation software we have found LCD codes with some parameters.

The program Generation indicates that there is no LCD code with parameters $[45,6,21]$, there exists a linear code $C_{[45,6,21]}$ with $1$ dimension hull.
Using the Generation software we find binary LCD codes with parameters $[41,6,19]$, $[43,7,19]$ and $[50,6,24]$.
The classification results can be found in \cite{Wang}.

\begin{table}[H]
    {\footnotesize
    \caption{Binary LCD codes constructed by expansion methods}
        \renewcommand\arraystretch{0.9}
        \begin{tabular}{ccc|ccc} \hline
        $C_{\mathbb{F}_4}$   & Transformation  &  $\Phi_n(C)$ & $C_{\mathbb{F}_4}$  & Transformation  &  $\Phi_n(C)$ \\  \hline\hline
        $[21,15,5]$  & id &  $[42,30,5]$ & $[21,18,3]$  & $\diag(w,w,w,1,...,1)$ &  $[42,36,3]$ \\
        $[23,15,6]$  & id &  $[46,30,6]$ & $[23,12,8]$  & id  &  $[46,24,8]$ \\
        $[24,9,12]$  & $\diag(w,w,w,1,...,1)$ & $[48,18,12]$ & $[24,11,10]$ & $\diag(w,1,1,1,...,1)$  & $[48,22,10]$ \\
        $[24,13,8]$  & $\diag(1,1,w,1,...,1)$  & $[48,26,8]$ & $[24,16,6]$  & $\diag(w,1,...,1)$ & $[48,32,6]$ \\
        $[25,11,11]$ & $\diag(1,w,1,...,1)$ & $[50,22,11]$ & $[25,10,12]$ & $\diag(1,w,w,1,...,1)$ & $[50,20,12]$ \\
        $[25,14,8]$  & $\diag(1,1,w,1,...,1)$ & $[50,28,8]$ & $[25,12,10]$ & id & $[50,24,10]$ \\
        $[25,17,6]$  & $\diag(w,w,1,...,1)$ & $[50,34,6]$ \\ \hline
\end{tabular}}
\begin{threeparttable}
    \begin{tablenotes}
        \footnotesize
        \item[1]
        In this table, the codes over $\mathbb{F}_4=\{0,1, w, w^2\}(w^2 = w + 1)$ are best known linear codes which can be found in codetables \cite{GTB22}.
      \end{tablenotes}
    \end{threeparttable}
\end{table}

\begin{table}[H]
    \caption{Binary LCD codes constructed by extension methods}
    {\footnotesize
        \renewcommand\arraystretch{0.9}
        \begin{tabular}{cccc} \hline
        $C$ & $\bm{x}$ or $\bm{y}$ &   $C(\bm{x})$ & References \\ \hline\hline
        $[42,36,3]$ & [001001]  & $[43,37,3]$ & Thm.~\ref{simpext} \\
        $[43,37,3]$ & [010001]  & $[44,38,3]$ & Thm.~\ref{simpext} \\
        $[44,38,3]$ & [000011]  & $[45,39,3]$ & Thm.~\ref{simpext} \\
        $[45,39,3]$ & [001010]  &  $[46,40,3]$ & Thm.~\ref{simpext} \\
        $[46,40,3]$ & [001011]  &  $[47,41,3]$ & Thm.~\ref{simpext} \\
        $[47,41,3]$ & [001100]  &  $[48,42,3]$ & Thm.~\ref{simpext} \\
        $[48,42,3]$ & [011111]  &  $[49,43,3]$ & Thm.~\ref{simpext} \\
        $[49,43,3]$ & [010110]  &  $[50,44,3]$ & Thm.~\ref{simpext} \\
        $[47,30,7]$ & [00101001110100000]  &  $[48,31,7]$ & Thm.~\ref{simpext} \\
        $[38,8,16]$ & [000000000000010110111111111110]  &  $[39,9,15]$ & Thm.~\ref{simpext} \\
        $[37,10,13]$ & [010001111101001010111110100]  &  $[38,11,13]$ & Thm.~\ref{simpext} \\
        $[39,14,11]$ & [0000001011111100010010011]  &  $[40,15,11]$ & Thm.~\ref{simpext} \\
        $[46,7,20]$ & [000000000000000011101111001111111110101]  &  $[47,8,19]$ & Thm.~\ref{simpext} \\
        $[48,8,20]$ & [0000000000000000010111111011111101010111]  &  $[49,9,19]$ & Thm.~\ref{simpext} \\
        $[45,18,12]$ & [000000000000001011101011111]  &  $[46,19,11]$ & Thm.~\ref{simpext} \\
        $[46,19,11]$ & [000000000100110101111111001]  &  $[47,20,11]$ & Thm.~\ref{simpext} \\
        $[47,20,11]$ & [000000101111101111011011110]  &  $[48,21,11]$ & Thm.~\ref{simpext} \\
        $[49,31,8]$ & [0...0000000000011001111]  &  $[50,32,7]$ & Thm.~\ref{thseff}\\
        $[40,9,16]$ & [0...00000000000001111111110111000101]  &  $[41,10,15]$ & Thm.~\ref{thseff} \\
        $[42,10,16]$ & [00000000010100010101101011110111]  &  $[43,11,15]$ & Thm.~\ref{simpext} \\
        $[45,6,21]$ & [000001] & $[46,6,21]$ & Cor.~\ref{hhbit} \\ \hline
\end{tabular}}
\begin{threeparttable}
\begin{tablenotes}
    \footnotesize
    \item[2]
    In this table, the even-like codes $C_{[49,31,8]}$, $C_{[40,9,16]}$ are extended from $C_{[48,31,7]}$, $C_{[39,9,15]}$ respectively,
    and $\dim(\Hull(C_{[49,31,8]}))=\dim(\Hull(C_{[40,9,16]}))=1$. The code $C_{[45,6,21]}$ is obtained by program Generation and $\dim(\Hull(C_{[45,6,21]}))=1$.
  \end{tablenotes}
\end{threeparttable}
\end{table}

\begin{table}[H]
    \caption{Binary LCD codes constructed by extending code length from $n$ to $n+1$ or $n+2$}
    {\footnotesize
        \renewcommand\arraystretch{0.9}
        \begin{tabular}{cc|cc|cc|cc} \hline
        $C$  &   $C'$  & $C$   &   $C'$ & $C$  &  $C'$ & $C$  &  $C'$ \\ \hline  \hline
        $[39,33,3]$   &  $[41,33,4]$ & $[40,34,3]$   &  $[41,34,4]$ & $[41,35,3]$   &  $[43,35,4]$ & $[42,36,3]$   &  $[43,36,4]$ \\

        $[43,37,3]$   &  $[45,37,4]$ & $[44,38,3]$   &  $[45,38,4]$ & $[45,39,3]$   &  $[47,39,4]$ & $[46,40,3]$   &  $[47,40,4]$ \\

        $[47,41,3]$   &  $[49,41,4]$ & $[48,42,3]$   &  $[49,42,4]$ & $[42,30,5]$   &  $[43,30,6]$ & $[43,31,5]$   &  $[45,31,6]$ \\
        $[44,32,5]$   &  $[45,32,6]$ & $[45,33,5]$   &  $[47,33,6]$ & $[46,34,5]$   &  $[47,34,6]$ & $[47,35,5]$   &  $[49,35,6]$ \\
        $[48,36,5]$   &  $[49,36,6]$ & $[41,24,7]$   &  $[42,24,8]$ & $[42,25,7]$   &  $[44,25,8]$ & $[43,26,7]$   &  $[44,26,8]$ \\
        $[44,27,7]$   &  $[46,27,8]$ & $[45,28,7]$   &  $[46,28,8]$ & $[46,29,7]$   &  $[48,29,8]$ & $[38,11,13]$  &  $[40,11,14]$ \\
        $[41,6,19]$   &  $[42,6,20]$ & $[46,6,21]$   &  $[47,6,22]$ & $[43,7,19]$   &  $[45,7,20]$ & $[48,7,21]$  &  $[50,7,22]$ \\
        $[47,8,19]$   &  $[48,8,20]$ & $[41,21,9]$   &  $[43,21,10]$ &$[46,10,17]$   &  $[47,10,18]$ & $[47,11,17]$   &  $[49,11,18]$ \\
        $[48,12,17]$  &  $[49,12,18]$ & $[41,10,15]$   &  $[42,10,16]$ & $[43,11,15]$  & $[45,11,16]$  & $[45,12,15]$   &  $[46,12,16]$ \\

        $[39,7,17]$   &  $[41,7,18]$ & $[43,17,11]$  &  $[45,17,12]$ & $[39,7,17]$   &  $[41,7,18]$ & $[44,18,11]$   &  $[45,18,12]$ \\
        $[43,15,13]$  &  $[45,15,14]$ & $[40,19,9]$  &  $[41,19,10]$ &  $[46,23,9]$  &  $[48,23,10]$ & $[47,24,9]$   &  $[48,24,10]$ \\
        $[48,25,9]$   &  $[50,25,10]$ & $[49,26,9]$  &  $[50,26,10]$ & $[46,19,11]$  &  $[48,19,12]$ & $[40,12,13]$   &  $[41,12,14]$ \\
        $[40,29,5]$   &  $[42,29,6]$ & $[41,13,13]$  &  $[43,13,14]$ & $[45,22,9]$   &  $[46,22,10]$ & $[47,20,11]$   &  $[48,20,12]$\\
        $[48,21,11]$  &  $[50,21,12]$ & $[42,9,17]$  &  $[44,9,18]$ & $[48,31,7]$  &  $[50,31,8]$ & $[39,14,11]$  &  $[40,14,12]$ \\ \hline
\end{tabular}}
\begin{threeparttable}
    \begin{tablenotes}
        \footnotesize
        \item[3] In this table, we use Lemma~\ref{pross3}(1), (2) to construct binary LCD codes.
        In fact, we can use Theorem~\ref{muccs} instead of Lemma~\ref{pross3}(2) to construct different LCD codes.
      \end{tablenotes}
    \end{threeparttable}
\end{table}

\begin{table}[H]
    \caption{Binary LCD codes constructed by shortening or puncturing methods}
    {  \footnotesize
        \renewcommand\arraystretch{0.9}
        \begin{tabular}{cccc|cccc} \hline
        $C$ & Coor. &  $C'$ & References& C & Coor. & $C'$ & References \\ \hline \hline
        $[42,36,3]$ & (2)  &  $[41,35,3]$ & Prop.~\ref{pro34} & $[41,35,3]$ & (1)  &  $[40,34,3]$ & Prop.~\ref{pro34} \\
        $[40,34,3]$ & (2)  &  $[39,33,3]$ & Prop.~\ref{pro34} & $[49,37,5]$ & (1)  &  $[48,36,5]$ & Prop.~\ref{pro34} \\
        $[48,36,5]$ & (5)  &  $[47,35,5]$ & Prop.~\ref{pro34} & $[47,35,5]$ & (1)  &  $[46,34,5]$ & Prop.~\ref{pro34} \\
        $[46,34,5]$ & (4)  &  $[45,33,5]$ & Prop.~\ref{pro34} & $[45,33,5]$ & (1)  &  $[44,32,5]$ & Prop.~\ref{pro34} \\
        $[44,32,5]$ & (3)  &  $[43,31,5]$ & Prop.~\ref{pro34} & $[43,31,5]$ & (1)  &  $[42,30,5]$ & Prop.~\ref{pro34} \\
        $[47,30,7]$ & (4)  &  $[46,29,7]$ & Prop.~\ref{pro34} & $[46,29,7]$ & (3)  &  $[45,28,7]$ & Prop.~\ref{pro34} \\
        $[45,28,7]$ & (2)  &  $[44,27,7]$ & Prop.~\ref{pro34} & $[44,27,7]$ & (1)  &  $[43,26,7]$ & Prop.~\ref{pro34} \\
        $[43,26,7]$ & (2)  &  $[42,25,7]$ & Prop.~\ref{pro34} & $[42,25,7]$ & (1)  &  $[41,24,7]$ & Prop.~\ref{pro34} \\

        $[50,27,9]$ & (1)  &  $[49,26,9]$ & Prop.~\ref{pro34} & $[49,26,9]$ & (3)  &  $[48,25,9]$ & Prop.~\ref{pro34} \\
        $[48,25,9]$ & (1)  &  $[47,24,9]$ & Prop.~\ref{pro34} & $[47,24,9]$ & (1)  &  $[46,23,9]$ & Prop.~\ref{pro34} \\
        $[46,23,9]$ & (2)  &  $[45,22,9]$ & Prop.~\ref{pro34} & $[50,14,17]$ & (1)  &  $[49,13,17]$ & Prop.~\ref{pro34} \\
        $[49,13,17]$ & (2)  &  $[48,12,17]$ & Prop.~\ref{pro34} & $[48,12,17]$ & (1)  &  $[47,11,17]$ & Prop.~\ref{pro34} \\
        $[47,11,17]$ & (1)  &  $[46,10,17]$ & Prop.~\ref{pro34} & $[46,13,15]$ & (1)  &  $[45,12,15]$ & Prop.~\ref{pro34} \\
        $[42,14,13]$ & (3)  &  $[41,13,13]$ & Prop.~\ref{pro34} & $[50,18,13]$ & (4)  &  $[49,17,13]$ & Prop.~\ref{pro34} \\

        $[51,18,14]$ & (1)  &  $[50,18,13]$ & Prop.~\ref{pro33} & $[51,14,18]$ & (1)  &  $[50,14,17]$ & Prop.~\ref{pro33}\\
        $[50,14,17]$ & (1)  &  $[49,14,16]$ & Prop.~\ref{pro34} & $[49,14,16]$ & (1)  &  $[48,14,15]$ & Prop.~\ref{pro34}\\

        $[51,27,10]$ & (1)  &  $[50,27,9]$ & Prop.~\ref{pro33} & $[49,13,17]$ & (1)  &  $[48,13,16]$ & Prop.~\ref{pro34} \\
        $[48,13,16]$ & (38,46)  &  $[46,13,15]$ & Prop.~\ref{pro33}  & $[45,16,14]$ & (1)  &  $[44,16,13]$ & Prop.~\ref{pro33}\\
        $[44,16,13]$ & (2)  &  $[43,16,12]$ &  Prop.~\ref{pro34} & $[43,16,12]$ & (1)  &  $[42,16,11]$ & Prop.~\ref{pro34} \\
        $[49,30,8]$ & (31)  &  $[48,30,8]$ & Prop.~\ref{pro33} & $[48,30,8]$ & (1)  &  $[47,30,7]$ & Prop.~\ref{pro33} \\
        $[50,6,24]$ & (1)  &  $[49,6,23]$ & Prop.~\ref{pro33} & $[43,9,18]$ & (1)  &  $[42,9,17]$ & Prop.~\ref{pro36} \\
        $[54,21,14]$ & (1,2,3)  &  $[51,18,14]$ & Prop.~\ref{pro35} \\ \hline
    \end{tabular}}
\begin{threeparttable}
\begin{tablenotes}
   \footnotesize
   \item[4] In this table, the codes $C_{[54,21,14]}$ and $C_{[43,9,18]}$ are best known linear codes obtained by Magma and $\dim(\Hull(C_{[54,21,4]}))=3$, $\dim(\Hull(C_{[43,9,18]}))=1$.
 \end{tablenotes}
\end{threeparttable}
\end{table}

\vskip 3mm
In Tables 5-6, we mark * those optimal LCD codes we have found and use boldface for those LCD codes we have found better than previous codes.
These improved codes were not previously known and cannot be derived using the inequality $d_{\LCD}(n,k) \le d_{\LCD}(n+1,k)$.
If the exact value of $d_{\LCD}(n,k)$ is not determined, we provide an interval.

\begin{table}[H]
    \centering
    {\footnotesize
    \renewcommand\arraystretch{1.1}
    \caption{Bounds on distances of binary LCD codes with $38 \le n \le 40$, $9 \le k \le 15$}
        \begin{tabular}{l|lllllllllll}\hline
            $n\backslash k$ & 9 & 10  & 11  &   12    & 13   &    14    &   15 \\ \hline \hline
            38 & 14-15 & 14    & \textbf{13}-14  & 12-14    & 11-12  & 10-12    & 10-12  \\
            39 & \textbf{15}-16 & 14-15 & 13-14  & 12-14    & 11-13  & 11-12    & 10-12   \\
            40 & 15-16 & 14-16 & \textbf{14}-15  & 13-14    & 12-14  & \textbf{12}-13    & \textbf{11}-12   \\\hline
    \end{tabular}}
\end{table}

\begin{table}[H]
    {\footnotesize
    \renewcommand\arraystretch{1.1}
    \caption{Bounds on distances of binary LCD codes with $41 \le n \le 50$, $6 \le k \le n-6$}
    \begin{tabular}{l|lllllllllll}\hline
    $n\backslash k$  & 6  & 7     & 8                  & 9               & 10              & 11              & 12              & 13             & 14             & 15 \\ \hline \hline
    41 & 19* & 18*             & 17*              & 16*             & \textbf{15}-16  & 14-16           & \textbf{14}-15  & \textbf{13}-14 & 12-14          & 11-13 \\
    42 & 20* & 18              & 18*              & 17*             & 16*             & 14-16           & 14-16           & 13-14          & \textbf{13}-14 & \textbf{12}-14 \\
    43 & 20  & 19*             & 18               & 17-18           & 16-17           & \textbf{15}-16  & 14-16           & \textbf{14}-15 & 14*            & \textbf{13}-14 \\
    44 & 20  & 19-20           & 18-19            & 18*             & 16-18           & 15-17           & 14-16           & 14-16          & 14-15          & 13-14 \\
    45 & 20  & 20*             & 18-20            & 18-19           & 16-18           & \textbf{16}-18  & \textbf{15}-16  & 14-16          & 14-16          & \textbf{14}-15 \\
    46 & 21* & 20              & 18-20            & 18-20           & \textbf{17}-19  & 16-18           & \textbf{16}-17  & \textbf{15}-16 & 14-16          & 14-16 \\
    47 & 22* & 20-21           & \textbf{19}-20   & 18-20           & \textbf{18}-20  & \textbf{17}-18  & 16-18           & 15-17          & 14-16          & 14-16 \\
    48 & 22  & \textbf{21}-22  & \textbf{20}-21   & 18-20           & 18-20           & 17-19           & \textbf{17}-18  & \textbf{16}-18 & \textbf{15}-17 & 14-16  \\
    49 & 23* & 21-22           & 20-22            & \textbf{19}-20  & 18-20           & \textbf{18}-20  & \textbf{18}-19  & \textbf{17}-18 & 16-18          & 14-17 \\
    50 & 24* & \textbf{22}-23  & 20-22            & 19-21           & 18-20           & 18-20           & 18-20           & 17-19          & 17-18          & 14-18 \\ \hline

    $n\backslash k$      & 16     & 17         & 18              & 19             & 20             & 21             & 22             & 23             & 24             & 25             \\ \hline\hline
    41    & 10-12           & 10-12          & 10-12           & \textbf{10}-11 & 10*            & \textbf{9}-10  & 8-9            & 8*             & \textbf{7}-8   & 6-8            \\
    42    & \textbf{11}-13  & 10-12          & 10-12           & 10-12          & 10             & 9-10           & 8-10           & 8-9            & 8*             & \textbf{7}-8   \\
    43    & \textbf{12}-14  & \textbf{11}-13 & 10-12           & 10-12          & 10-11          & 10*            & 8-10           & 8-10           & 8-9            & 7-8            \\
    44    & \textbf{13}-14  & 11-14          & \textbf{11}-12  & 10-12          & 10-12          & 10-11          & 8-10           & 8-10           & 8-10           & \textbf{8}-9   \\
    45    & \textbf{14}     & \textbf{12}-14 & \textbf{12}-13  & 10-12          & 10-12          & 10-12          & \textbf{9}-11  & 8-10           & 8-10           & 8-10           \\
    46    & 14-15           & 12-14          & 12-14           & 11-13          & 10-12          & 10-12          & \textbf{10}-12 & \textbf{9}-11  & 8-10           & 8-10           \\
    47    & 14-16           & 12-15          & 12-14           & \textbf{11}-14 & \textbf{11}-13 & 10-12          & 10-12          & 9-12           & \textbf{9}-11  & 8-10           \\
    48    & 14-16           & 12-16          & 12-15           & \textbf{12}-14 & \textbf{12}-14 & \textbf{11}-13 & 10-12          & \textbf{10}-12 & \textbf{10}-12 & \textbf{9}-12  \\
    49    & 14-16           & \textbf{13}-16 & 12-15           & 12-14          & 12-14          & 11-14          & 10-13          & 10-12          & 10-12          & 9-12           \\
    50    & 14-17           & 13-16          & \textbf{13}-16  & 12-15          & 12-14          & \textbf{12}-14 & \textbf{11}-14 & 10-13          & 10-12          & \textbf{10}-12 \\ \hline
    $n\backslash k$ & 26  & 27   & 28   & 29   & 30   & 31   & 32  & 33   & 34   & 35            \\ \hline\hline
    41   & 6-7            & 6    & 6    & 5-6  & 4-5  & 4        & 4   & 4* & 4*  & 3*    \\
    42   & 6-8            & 6-7  & 6    & 6*   & \textbf{5}-6 & 4-5  & 4     & 4 & 4 & 3-4       \\
    43   & \textbf{7}-8   & 6-8  & 6-7  & 6    & 6*   & \textbf{5}-6 & 4-5    & 4 & 4  & 4*           \\
    44   & 8*             & \textbf{7}-8 & 6-8  & 6-7  & 6    & 5-6  & \textbf{5}-6   & 4-5  & 4 & 4           \\
    45   & 8-9            & 7-8  & \textbf{7}-8 & 6-8  & 6-7  & 6*   & 6*    & \textbf{5}-6 & 4-5 & 4         \\
    46   & 8-10           & \textbf{8}-9 & 8*   & \textbf{7}-8 & 6-8  & 6-7    & 6    & 5-6  & \textbf{5}-6 & 4-5         \\
    47   & 8-10           & 8-10 & 8-9  & 7-8  & \textbf{7}-8 & 6-8  & 6-7     & 6* & 6*  & \textbf{5}-6       \\
    48   & 8-10           & 8-10 & 8-10 & \textbf{8}-9 & 8*   & \textbf{7}-8  & 6-8    & 6-7 & 6  & 5-6      \\
    49   & \textbf{9}-11  & 8-10 & 8-10 & 8-10 & 8    & 7-8  & 6-8    & 6-8 & 6-7 & 6*        \\
    50   & \textbf{10}-12 & \textbf{9}-10 & 8-10 & 8-10 & 8-9  & 8  & \textbf{7}-8  & 6-8 & 6-8  & 6-7          \\ \hline
    $n\backslash k$ & 36 & 37 &  38  & 39 & 40 & 41 & 42 & 43 & 44  \\ \hline \hline
    42  & 3* &  &    &     &    &          \\
    43  & 4*  & 3* &  &   &    &    &       \\
    44  & 4 & 3-4  & 3*    &      \\
    45  & 4  & 4* & 4* &   3*   &      \\
    46  & 4 & 4  & 4   & 3-4 &   3*  \\
    47  & 4-5 & 4 & 4    & 4*  & 4* & 3*\\
    48  & \textbf{5}-6 & 4-5 & 4    & 4   & 4 & 3-4  & 3* \\
    49  & 6* & \textbf{5}-6 & 4-5  & 4    & 4 & 4*   & 4*  & 3*  \\
    50  & 6 & 5-6 & \textbf{5}-6 & 4    & 4  & 4 & 4  & 3-4  &  3*  \\ \hline
    \end{tabular}}
\end{table}

\section{Conclusion}

In this paper, we have enriched the methods for constructing LCD codes from linear codes or LCD codes,
such as expanding, extending, puncturing codes first and then extending.
We provide a general lower bound for the distance of binary LCD codes using expanding methods,
and also demonstrate that this lower bound is tight.
In addition, we consider extending general linear codes to construct LCD codes,
which is a practical approach. Finally, we utilize some known methods, in combination with our own methods and computer-aided software,
to construct or discover new LCD codes with favorable parameters. In particular, we improve some previously known range of $d_{\LCD}(n, k)$ of lengths $38 \le n \le 40$ and dimensions $9 \le k \le 15$.
For lengths $41 \le n \le 50$ and dimensions $6 \le k \le n-6$,
we present novel values or ranges of $d_{\LCD}(n, k)$ that have not been discussed previously.

\vskip 2mm
\noindent\textbf{Acknowledgement.}
This work was supported by NSFC (Grant Nos. 11871025, 12271199 and 12171191).

\end{document}